\documentclass[12pt]{article}
\usepackage{amssymb}
\usepackage{graphics}
\usepackage{graphicx}
\begin{document}



\def\a{\alpha}
\def\b{\beta}
\def\d{\delta}
\def\e{\epsilon}
\def\g{\gamma}
\def\h{\mathfrak{h}}
\def\k{\kappa}
\def\l{\lambda}
\def\o{\omega}
\def\p{\wp}
\def\r{\rho}
\def\t{\theta}
\def\s{\sigma}
\def\z{\zeta}
\def\x{\xi}
 \def\A{{\cal{A}}}
 \def\B{{\cal{B}}}
 \def\C{{\cal{C}}}
 \def\D{{\cal{D}}}
\def\G{\Gamma}
\def\K{{\cal{K}}}
\def\O{\Omega}
\def\R{\bar{R}}
\def\T{{\cal{T}}}
\def\L{\Lambda}
\def\f{E_{\tau,\eta}(sl_2)}
\def\E{E_{\tau,\eta}(sl_n)}
\def\Zb{\mathbb{Z}}
\def\Cb{\mathbb{C}}

\def\R{\overline{R}}

\def\beq{\begin{equation}}
\def\eeq{\end{equation}}
\def\bea{\begin{eqnarray}}
\def\eea{\end{eqnarray}}
\def\ba{\begin{array}}
\def\ea{\end{array}}
\def\no{\nonumber}
\def\le{\langle}
\def\re{\rangle}
\def\lt{\left}
\def\rt{\right}

\newtheorem{Theorem}{Theorem}
\newtheorem{Definition}{Definition}
\newtheorem{Proposition}{Proposition}
\newtheorem{Lemma}{Lemma}
\newtheorem{Corollary}{Corollary}
\newcommand{\proof}[1]{{\bf Proof. }
        #1\begin{flushright}$\Box$\end{flushright}}

\baselineskip=20pt

\newfont{\elevenmib}{cmmib10 scaled\magstep1}
\newcommand{\preprint}{
   \begin{flushleft}
   \end{flushleft}\vspace{-1.3cm}
   \begin{flushright}\normalsize
   \end{flushright}}
\newcommand{\Title}[1]{{\baselineskip=26pt
   \begin{center} \Large \bf #1 \\ \ \\ \end{center}}}
\newcommand{\Author}{\begin{center}
   \large \bf
Wen-Li Yang${}^{a}$, ~Xi Chen${}^{a}$, ~Jun Feng${}^{a}$,~Kun Hao${}^{a}$,~Kang-Jie Shi ${}^a$,
~Cheng-Yi Sun${}^a$,~Zhan-Ying Yang${}^b$ ~and~Yao-Zhong Zhang ${}^c$
 \end{center}}
\newcommand{\Address}{\begin{center}

     ${}^a$ Institute of Modern Physics, Northwest University,
     Xian 710069, P.R. China\\
     ${}^b$ The Department of Physics, Northwest University,
     Xian 710069, P.R. China \\
     ${}^c$ The University of Queensland, School of Mathematics and Physics,  Brisbane, QLD 4072,
     Australia\\
   \end{center}}
\newcommand{\Accepted}[1]{\begin{center}
   {\large \sf #1}\\ \vspace{1mm}{\small \sf Accepted for Publication}
   \end{center}}

\preprint
\thispagestyle{empty}
\bigskip\bigskip\bigskip

\Title{Domain wall partition function of the eight-vertex model with
a non-diagonal reflecting end} \Author

\Address
\vspace{1cm}

\begin{abstract}
With the help of the Drinfeld twist or factorizing
F-matrix for the  eight-vertex SOS model, we obtain the explicit
determinant expression of the partition function
of the eight-vertex model  with a generic
non-diagonal reflecting end and domain wall boundary condition. Our
result shows that, contrary to the eight-vertex model without a
reflection end, the partition function can be expressed as a single determinant.

\vspace{1truecm} \noindent {\it PACS:} 03.65.Fd; 04.20.Jb;
05.30.-d; 75.10.Jm

\end{abstract}
\newpage
\section{Introduction}
\label{intro} \setcounter{equation}{0}
The domain wall (DW) boundary condition of a statistical model on a finite two-dimensional
lattice was first introduced in \cite{Kor82} for the six-vertex model. The partition function of the model (or
DW partition function) was then expressed in terms of a determinant \cite{Ize87,Ize92}. Such a
determinant representation of the partition function has played an
important role in constructing norms of Bethe states, correction
functions \cite{Ess95,Kor93,Kit99} and  thermodynamical properties
of the six-vertex model \cite{Ble05}, and also in analyzing the  Toda
theories \cite{Sog93}. Moreover, it has been proven to be very
useful in solving some pure mathematical problems, such as the
problem of alternating sign matrices \cite{Kup96}. Recently, the determinant representations of the DW partition
function have been obtained for various models \cite{Car06,Fod08,Zha07,Pak08,Yan09-1,Ros09,Hao10}.

For a two-dimensional statistical model with a reflection end \cite{Tsu98}, in addition to the local interaction vertex,
a reflecting matrix or K-matrix \cite{Skl88} which describes the boundary interactions needs to be introduced
at the reflection end of the lattice (see figure 4 below). The DW partition function of the six-vertex model with a diagonal reflection
end was exactly calculated and expressed in terms some determinant \cite{Tsu98}.
Such determinant representation of the model was re-derived
\cite{Wan02, Kit07} by using  the Drinfeld twist  or factorizing F-matrix \cite{Dri83} method \cite{Mai00}.
However, it is highly non-trivial to generalize this result to models  with non-diagonal boundary terms
\cite{Nep04,Nep03,Cao03,Yan04-1,Gie05,Yan04,Gal05,
Gie05-1,Baj06,Yan05,Doi06,Mur06,Bas07,Yan06,Gal08,Yan07,Ami10,Cra10}.
Very recently, determinant representations of DW partition functions have been obtained for the trigonometric
SOS model with reflection end \cite{Fil10} and the six-vertex model with a non-diagonal reflection end \cite{Yan11}.

Among solvable models, elliptic ones stand out as a particularly
important class due to the fact that most trigonometric and
rational models can be obtained from them by  certain limits. In
this paper, we focus on the most fundamental elliptic model---the
eight-vertex model \cite{Bax82} whose trigonometric limit
gives the six-vertex model. Here we obtain the determinant representation of  the DW partition
function of the eight-vertex model with a non-diagonal reflection end specified by the generic non-diagonal
K-matrix \cite{Ina94,Hou95}. Our result shows that contrary to the eight-vertex model without reflection end
whose DW partition function can only be expressed as  a sum of determinants \cite{Pak08,Yan09-1,Ros09}, the DW partition function of the eight-vertex
model with a non-diagonal reflection end  can be
represented as a single determinant. Such a single determinant expression will be essential for the study
of the Bethe vectors of the open XYZ chain with non-diagonal boundary terms.

The paper is organized as follows.  In section 2, after introducing our notation and some basic ingredients,
we construct the four boundary states which specify the DW boundary condition of the eight-vertex model
with a non-diagonal reflection end. In section 3, using the vertex-face correspondence relation we express the
DW boundary partition function in terms of the matrix element of the product of the (face type)
pseudo-particle creation operators. In section 4, with help of the F-matrix of the eight-vertex SOS model we give  the completely
symmetric and polarization free representations of the
pseudo-particle creation operators  in the F-basis. In section 5,
we obtain the determinant representation of
the DW partition function.  In section 6, we summarize our results and give some discussions.


\section{ Eight-vertex model with a reflecting end}
\label{XXZ} \setcounter{equation}{0}
In this section, we briefly review the DW boundary condition for
the eight-vertex model  with non-diagonal reflecting end on an $N\times 2N$ rectangular lattice.

\subsection{The eight-vertex R-matrix and associated K-matrix}

Let us fix $\tau$ such that ${\rm Im}(\tau)>0$ and $\Lambda_{\tau}$ be the lattice generated by
$1$ and $\tau$. Introduce the following elliptic functions
\bea
\theta\lt[\begin{array}{c} a\\b
  \end{array}\rt](u,\tau)&=&\sum_{n=-\infty}^{\infty}
  \exp\lt\{i\pi\lt[(n+a)^2\tau+2(n+a)(u+b)\rt]\rt\},\label{Function-a-b}\\
\theta^{(j)}(u)&=&\theta\lt[\begin{array}{c}\frac{1}{2}-\frac{j}{2}\\
 [2pt]\frac{1}{2}
 \end{array}\rt](u,2\tau),\quad j=1,2;\qquad
 \s(u)=\theta\lt[\begin{array}{c}\frac{1}{2}\\[2pt]\frac{1}{2}
 \end{array}\rt](u,\tau).
 \label{Function-j}\eea The
$\s$-function\footnote{Our $\s$-function is the
$\vartheta$-function $\vartheta_1(u)$ \cite{Whi50}. It has the
following relation with the {\it Weierstrassian\/} $\s$-function
$\s_w(u)$: $\s_w(u)\propto e^{\eta_1u^2}\s(u)$ with
$\eta_1=\pi^2(\frac{1}{6}-4\sum_{n=1}^{\infty}\frac{nq^{2n}}{1-q^{2n}})
$ and $q=e^{i\tau}$.}
 satisfies the so-called Riemann
identity:\bea
&&\s(u+x)\s(u-x)\s(v+y)\s(v-y)-\s(u+y)\s(u-y)\s(v+x)\s(v-x)\no\\
&&~~~~~~=\s(u+v)\s(u-v)\s(x+y)\s(x-y),\label{identity}\eea which
will be useful in the following. Moreover, for any $\a=(\a_1,\a_2),\,\a_1,\a_2\in\Zb_2$, we
can introduce a function $\s_{\a}(u)$ as follow
\bea
  \s_{\a}(u)=\theta\lt[\begin{array}{c}\frac{1}{2}+\frac{\a_1}{2}\\[2pt]\frac{1}{2}+\frac{\a_2}{2}
 \end{array}\rt](u,\tau),\quad\quad \a_1,\a_2\in \Zb_2.\label{sigma-function}
\eea The above definition implies the identification
  $\s_{(0,0)}(u)=\s(u)$.

Let $V$ be a two-dimensional vector space $\Cb^2$ and
$\{\e_i|i=1,2\}$ be the orthonormal basis  of $V$ such that
$\langle \e_i,\e_j\rangle=\d_{ij}$. The well-known eight-vertex
model R-matrix $\R(u)\in {\rm End}(V\otimes V)$ is given by \bea
\R(u)=\lt(\begin{array}{llll}a(u)&&&d(u)\\&b(u)&c(u)&\\
&c(u)&b(u)&\\d(u)&&&a(u)\end{array}\rt). \label{r-matrix}\eea The
non-vanishing matrix elements  are \cite{Bax82}\bea
&&a(u)=\frac{\t^{(1)}(u)\,\t^{(0)}(u+\eta)\,\s(\eta)}
{\t^{(1)}(0)\, \t^{(0)}(\eta)\,\s(u+\eta)},\quad
b(u)=\frac{\t^{(0)}(u)\,
 \t^{(1)}(u+\eta)\,\s(\eta)}
{\t^{(1)}(0)\,\t^{(0)}(\eta)\,\s(u+\eta)},\no\\[6pt]
&&c(u)=\frac{\t^{(1)}(u)\,
 \t^{(1)}(u+\eta)\,\s(\eta)}
{\t^{(1)}(0)\, \t^{(1)}(\eta)\,\s(u+\eta)},\quad
d(u)=\frac{\t^{(0)}(u)\,
 \t^{(0)}(u+\eta)\,\s(\eta)}
{\t^{(1)}(0)\t^{(1)}(\eta)\,\s(u+\eta)}.\label{r-func}\eea Here
$u$ is the spectral parameter and  $\eta$ is the so-called crossing
parameter. The R-matrix satisfies the quantum Yang-Baxter equation
(QYBE)
\bea \R_{1,2}(u_1-u_2)\R_{1,3}(u_1-u_3)\R_{2,3}(u_2-u_3)
=\R_{2,3}(u_2-u_3)\R_{1,3}(u_1-u_3)\R_{1,2}(u_1-u_2).\label{QYBE}\eea
Throughout we adopt the standard notation: for any
matrix $A\in {\rm End}(V)$, $A_j$ (or  $A^j$)is an embedding operator in the
tensor space $V\otimes V\otimes\cdots$, which acts as $A$ on the
$j$-th space and as identity on the other factor spaces;
$R_{i,j}(u)$ is an embedding operator of R-matrix in the tensor
space, which acts as identity on the factor spaces except for the
$i$-th and $j$-th ones.

For  a model with reflection end \cite{Skl88}, in addition to the R-matrix, one
needs to introduce  K-matrix $K(u)$ which
satisfies the reflection equation \cite{Che84} (RE)
 \bea &&\R_{1,2}(u_1-u_2)K_1(u_1)\R_{2,1}(u_1+u_2)K_2(u_2)\no\\
 &&~~~~~~=
K_2(u_2)\R_{1,2}(u_1+u_2)K_1(u_1)\R_{2,1}(u_1-u_2).\label{RE-V}\eea

In this paper, we consider the K-matrix $K(u)$ which is a
generic solution \cite{Ina94,Hou95} to the RE (\ref{RE-V}) associated with the R-matrix
(\ref{r-matrix}),
\bea K(u)=k_0(u)+k_x(u)\,\s^x+k_y\,\s^y+k_z(u)\,\s^z,\label{K-matrix}\eea
where $\s^x,\s^y,\s^z$ are the Pauli matrices and the coefficient functions are
\bea
&& k_0(u)=\frac{\s(2u)\,\s(\l_1+\l_2-\frac{1}{2})\,\s(\l_1+\zeta)\,\s(\l_2+\zeta)}
   {2\,\s(u)\,\s(-u+\l_1+\l_2-\frac{1}{2})\,\s(\l_1+\zeta+u)\,\s(\l_2+\zeta+u)},\no\\[6pt]
&& k_x(u)=\frac{\s(2u)\,\s_{(1,0)}(\l_1+\l_2-\frac{1}{2})\,\s_{(1,0)}(\l_1+\zeta)\,\s_{(1,0)}(\l_2+\zeta)}
   {2\,\s_{(1,0)}(u)\,\s(-u+\l_1+\l_2-\frac{1}{2})\,\s(\l_1+\zeta+u)\,\s(\l_2+\zeta+u)},\no\\[6pt]
&& k_y(u)=\frac{i\,\s(2u)\,\s_{(1,1)}(\l_1+\l_2-\frac{1}{2})\,\s_{(1,1)}(\l_1+\zeta)\,\s_{(1,1)}(\l_2+\zeta)}
   {2\,\s_{(1,1)}(u)\,\s(-u+\l_1+\l_2-\frac{1}{2})\,\s(\l_1+\zeta+u)\,\s(\l_2+\zeta+u)},\no\\[6pt]
&& k_z(u)=\frac{\s(2u)\,\s_{(0,1)}(\l_1+\l_2-\frac{1}{2})\,\s_{(0,1)}(\l_1+\zeta)\,\s_{(0,1)}(\l_2+\zeta)}
   {2\,\s_{(0,1)}(u)\,\s(-u+\l_1+\l_2-\frac{1}{2})\,\s(\l_1+\zeta+u)\,\s(\l_2+\zeta+u)}.\label{K-matrix-2-1} \eea
Here there are 3 free boundary parameters $\{\l_1,\l_2,\zeta\}$, which are related to different boundary interactions.
It is very convenient to introduce a vector $\l\in V$ associated with
the boundary parameters $\{\l_i\}$, \bea
 \l=\sum_{k=1}^2\l_k\e_k. \label{boundary-vector}
\eea


\subsection{The model}
The partition function of a statistical model on a two-dimensional
lattice  is defined by
\begin{eqnarray}
 Z=\sum \exp\{-\frac{E}{kT}\}, \nonumber
\end{eqnarray}
where $E$ is the energy of the system, $k$ is the Boltzmann
constant, $T$ is the temperature of the system,  and the summation
is taken over all possible configurations under certain
boundary condition such as the DW boundary condition. The model we
consider here has eight allowed bulk vertex configurations

\begin{center}
\includegraphics[width=\textwidth]{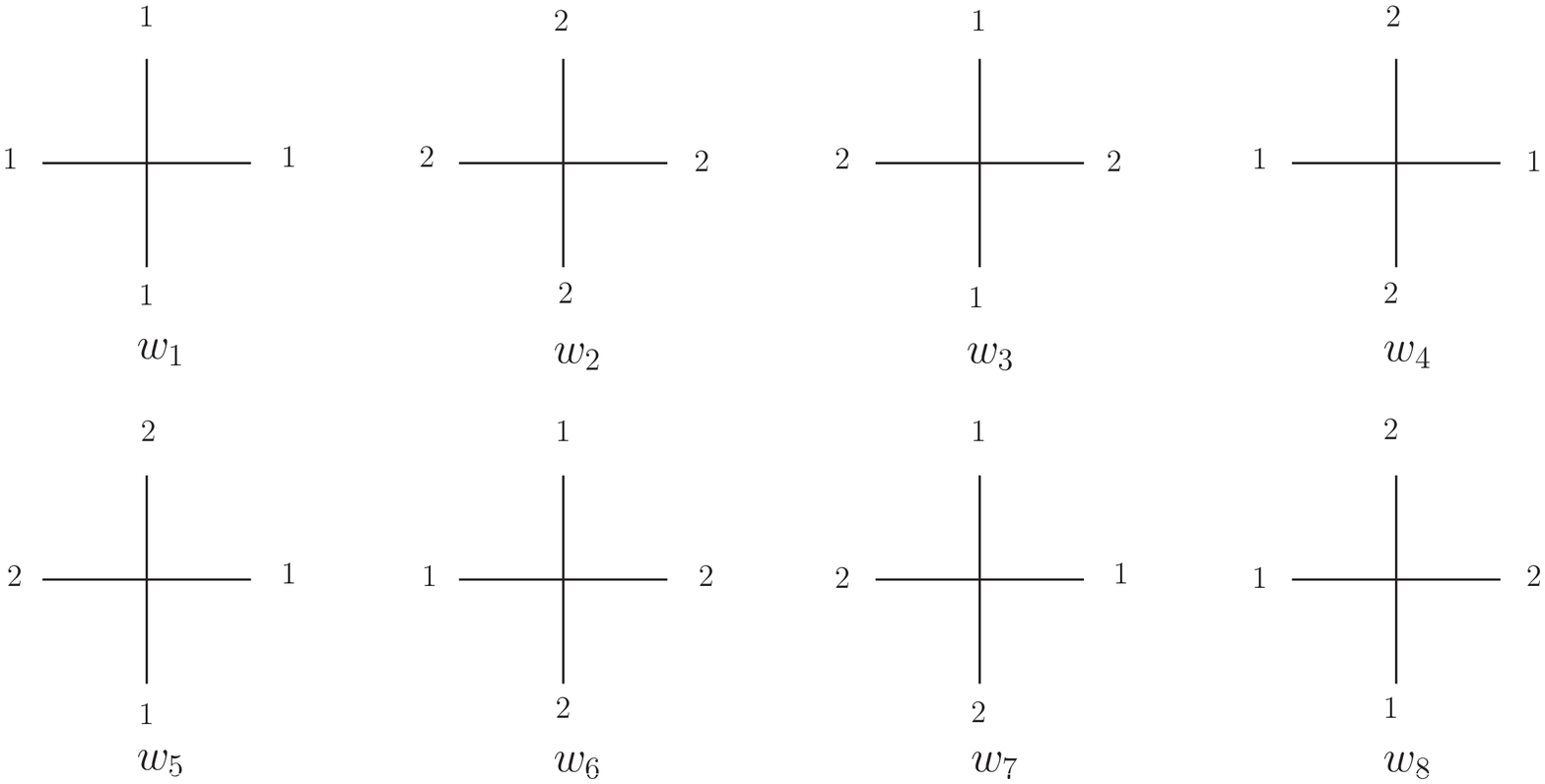}\\
{\small{\bf Figure 1.} Vertex configurations and
their associated Boltzmann weights.}
\end{center}


\noindent and four allowed configurations at each
reflection end


\begin{center}
\includegraphics[width=0.7\textwidth]{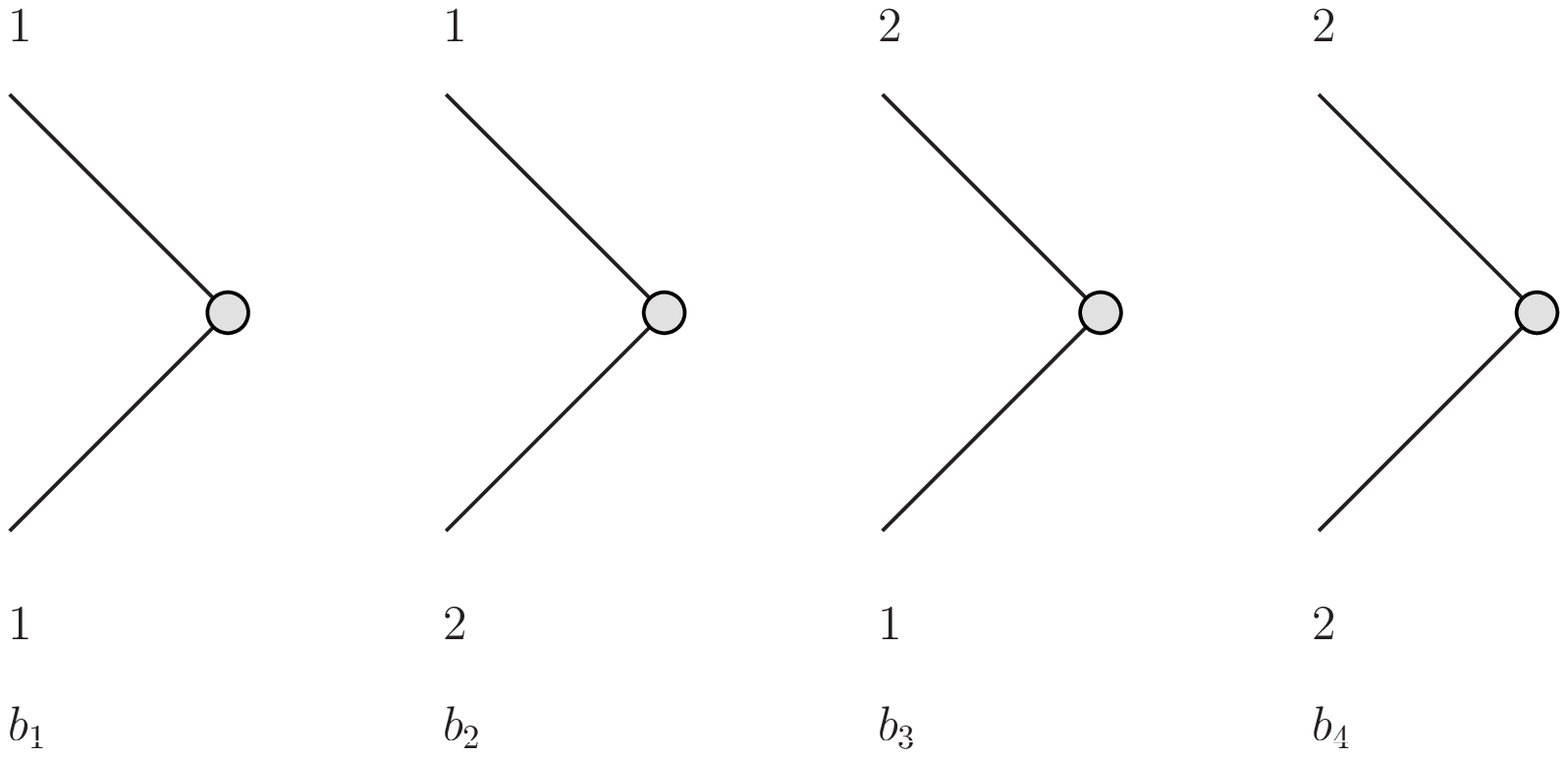}\\
{\small{\bf Figure 2.} Reflection ends and
the associated Boltzmann weights. }
\end{center}


\noindent where $1$ and $2$ respectively denote the spin up and
down states. Each of the eight bulk configurations is assigned a
statistical weight (or Boltzmann weight) $w_i$, while each of the
four reflection configurations is assigned a  weight $b_i$.
Then the partition function  of the model with a reflection end can be rewritten as
\begin{eqnarray}
Z=\sum\,{w_1}^{n_1}\,{w_2}^{n_2}\,{w_3}^{n_3}\,{w_4}^{n_4}\,{w_5}^{n_5}
 \,{w_6}^{n_6}\,{w_7}^{n_7}\,{w_8}^{n_8}\,b_1^{l_1}\,b_2^{l_2}\,b_3^{l_3}\,b_4^{l_4},\no
\end{eqnarray} where the summation is over all possible configurations with $n_i$
and $l_j$ being the number of vertices of type $i$ and the number of reflection ends of
type $j$ respectively. The bulk
Boltzmann weights  which we consider here have $Z_2$-symmetry, i.e.,
\begin{eqnarray}
  a\equiv w_1=w_2,\quad b\equiv w_3=w_4, \quad c\equiv
  w_5=w_6,\quad d\equiv w_7=w_8,
\end{eqnarray} and variables $a,b,c,d$ satisfy a function
relation, or equivalently, the local Boltzmann weights $\{w_i\}$
can be parameterized by the matrix elements of the eight-vertex
R-matrix (\ref{r-matrix}) as in  figure 3. At the same time, the weights
$\{b_i\}$ corresponding to the reflection end can be parameterized by the matrix
elements of the corresponding K-matrix (\ref{K-matrix})  as in figure 3.


\begin{center}
\includegraphics[width=0.8\textwidth]{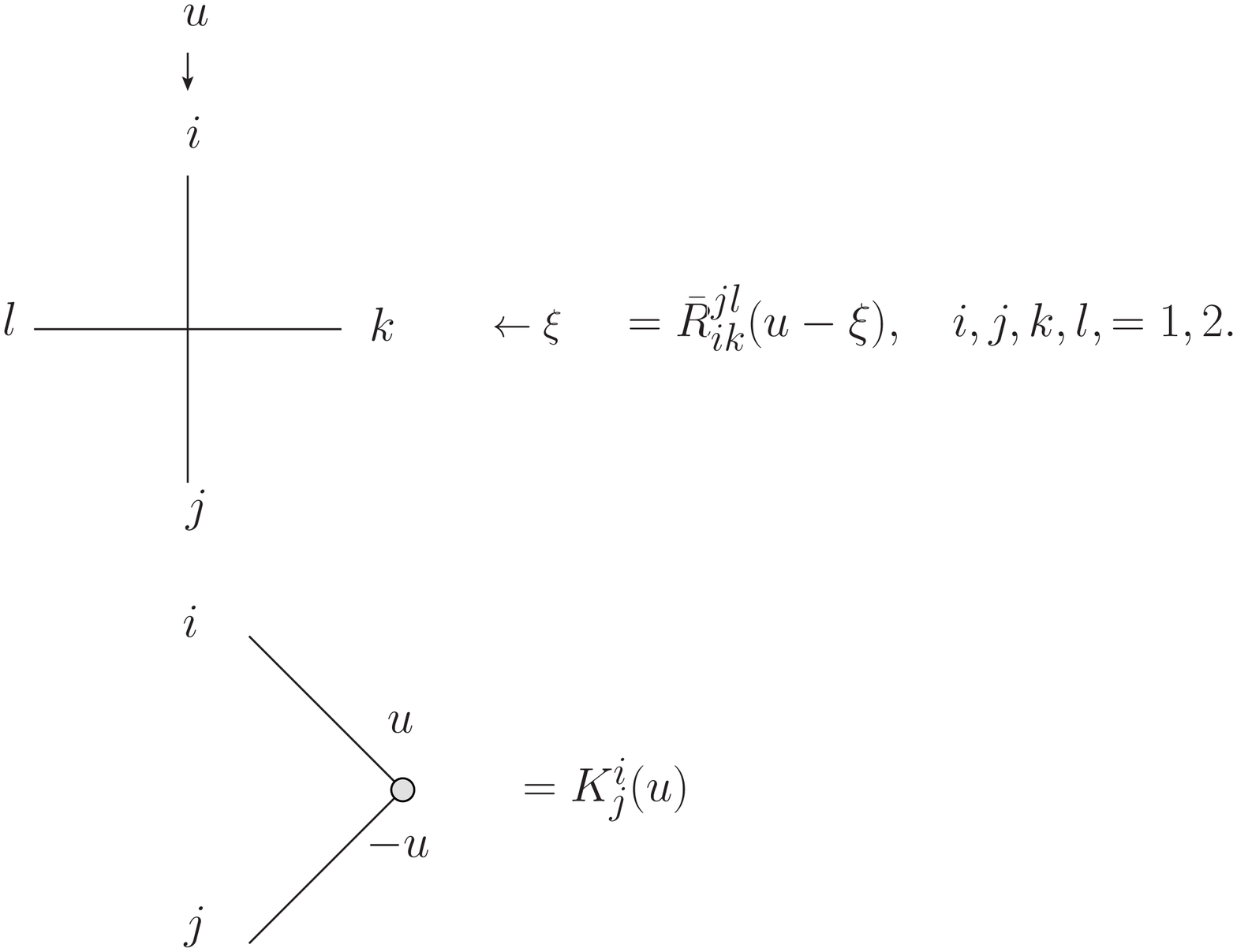}\\
{\small{\bf Figure 3.} The Boltzmann weights and elements of the eight-vertex R-matrix and K-matrix. }
\end{center}


\noindent Then the corresponding model is called the eight-vertex
model with a reflection end. Its
partition function is thus given by
\begin{eqnarray}
Z=\sum\,{a}^{n_1+n_2}\,{b}^{n_3+n_4}\,{c}^{n_5+n_6}\, {d}^{n_7+n_8}\,
 b_1^{l_1}\,b_2^{l_2}\,b_3^{l_3}\,b_4^{l_4}.\no
\end{eqnarray}

As shown in figure 3, to parameterize the bulk Boltzmann weights in terms of
the elements of the R-matrix, one needs to assign spectral
parameters $u$ and $\xi$ respectively to the vertical and
horizontal lines of each vertex of the lattice. In an inhomogeneous model, the statistical weights are
site-dependent. Hence two sets of spectral parameters
$\{u_{\alpha}\}$ and $\{\xi_i\}$ are needed, see figure 4. The
horizontal lines are enumerated by indices $1,\ldots,N$ with
spectral parameters $\{\xi_i\}$, while the vertical lines are
enumerated by indices $\bar{1},\ldots,\bar{N}$ with spectral
parameters $\{\bar{u}_{\alpha}\}$ (The $2N$ parameters $\{\bar{u}_{\alpha}\}$
are assigned as follow: $\bar{u}_{2i}=u_i$ and $\bar{u}_{2i+1}=-u_i$, as shown in figure 4.).
The DW boundary condition is
specified by four boundary states $|\Omega^{(2)}(\lambda)\rangle$,
$|\bar{\Omega}^{(1)}(\lambda)\rangle$,
$\langle\Omega^{(1)}(\lambda)|$ and
$\langle\bar{\Omega}^{(2)}(\lambda)|$ (the
definitions of the boundary states  will be given later,
see (\ref{Boundary-state-1})-(\ref{Boundary-state-4}) below).
These four states correspond to the particular choices of spin
states on the four boundaries of the lattice .

\vskip20mm
\begin{center}
\includegraphics[width=0.8\textwidth]{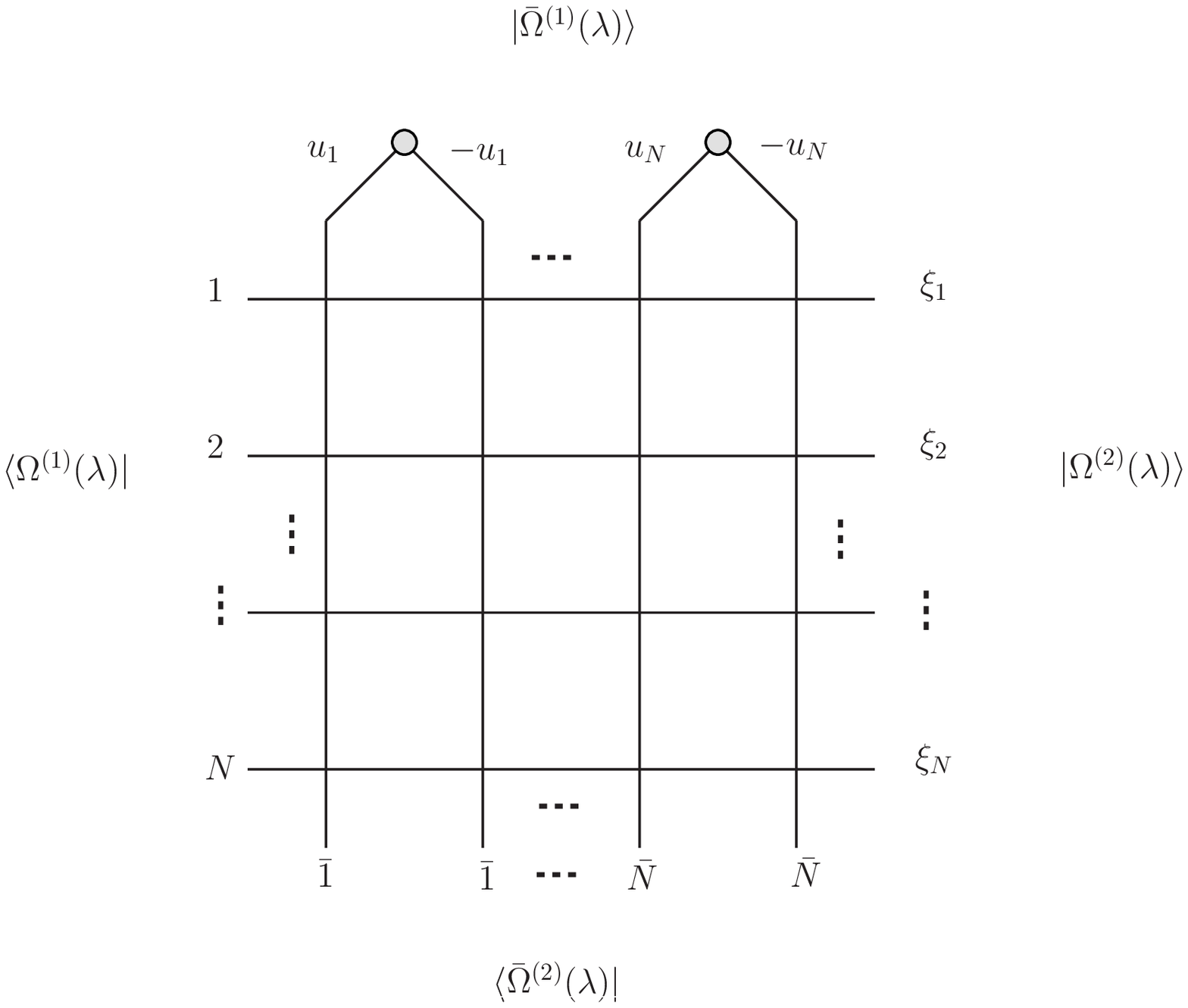}\\
{\small{\bf Figure 4.} The eight-vertex model with a non-diagonal reflection end and the DW condition.}
\end{center}

Some remarks are in order. The boundary states not only depend
on the spectral parameters ($|\Omega^{(2)}(\lambda)\rangle$ and
$\langle\Omega^{(1)}(\lambda)|$ depend on
$\{\xi_i\}$, while $|\bar{\Omega}^{(1)}(\lambda)\rangle$ and
$\langle\bar{\Omega}^{(2)}(\lambda)|$ depend on
$\{u_{\alpha}\}$) but also on two continuous parameters
$\lambda_1$ and $\lambda_2$.
However, in the
trigonometric limit (i.e., setting $\l_2=\frac{\tau}{2}$ and then
taking $\tau\rightarrow +i\infty$), the corresponding boundary
states $|\Omega^{(1)}(\lambda)\rangle$ and
$\langle\bar{\Omega}^{(1)}( \lambda)|$ (or
$|\bar{\Omega}^{(2)}(\lambda)\rangle$ and
$\langle\Omega^{(2)}(\lambda)|$) become the state of
all spin up and its dual (or the state of all spin down and its
dual) up to some over-all scalar factors.

The partition function of the eight-vertex model with a non-diagonal
reflection end specified by the generic K-matrix $K(u)$ (\ref{K-matrix})
and the DW boundary condition is a function of $2N+3$ variables $\{u_{\a}\}$,
$\{\xi_i\}$, $\l_1$, $\l_2$ and $\zeta$, which is denoted by
$Z_N(\{u_{\a}\};\{\xi_i\};\l;\zeta)$. Due to the fact that the local
Boltzmann weights of each vertex and reflection end of the lattice
are given by the matrix elements of the eight-vertex R-matrix and the associated K-matrix (see figure 3),
the partition function can be expressed in terms of the product of the
R-matrices, the K-matrices and the four boundary states
\bea
 Z_N(\{u_{\a}\};\{\xi_i\};\l;\zeta)
    \hspace{-0.2truecm}&=&\hspace{-0.2truecm}\langle\Omega^{(1)}(\l)
    |\langle\bar{\Omega}^{(2)}(\l)|\no\\
 &&\times \R_{\bar{1},N}(u_1\hspace{-0.1truecm}-\hspace{-0.1truecm}\xi_N)\ldots
    \R_{\bar{1},1}(u_1\hspace{-0.1truecm}-\hspace{-0.1truecm}\xi_1)K_{\bar{1}}(u_1)
    \R_{1,\bar{1}}(u_1\hspace{-0.1truecm}+\hspace{-0.1truecm}\xi_1)\ldots
    \R_{N,\bar{1}}(u_1\hspace{-0.1truecm}+\hspace{-0.1truecm}\xi_N)\no\\
 &&\qquad\qquad\vdots\no\\
 &&\times \R_{\bar{N},N}(u_N\hspace{-0.1truecm}-\hspace{-0.1truecm}\xi_N)\ldots
    \R_{\bar{N},1}(u_N\hspace{-0.1truecm}-\hspace{-0.1truecm}\xi_1)K_{\bar{N}}(u_N)
    \R_{1,\bar{N}}(u_N\hspace{-0.1truecm}+\hspace{-0.1truecm}\xi_1)\ldots
    \R_{N,\bar{N}}(u_N\hspace{-0.1truecm}+\hspace{-0.1truecm}\xi_N)\no\\[8pt]
 && \times |\bar{\Omega}^{(1)}(\l)\rangle
    |\Omega^{(2)}(\l)\rangle.\label{PF}\eea
One can rearrange the product of the R-matrices in (\ref{PF}) in
terms of  a product  of the so-called double-row monodromy matrices
\bea
 \hspace{-1truecm}Z_N(\{u_{\a}\};\{\xi_i\};\l;\zeta)
 \hspace{-0.38truecm}&=&\hspace{-0.38truecm}\langle\Omega^{(1)}(\l)
    |\langle\bar{\Omega}^{(2)}(\l)|\,
    \mathbb{T}_{\bar{1}}(u_1)\ldots \mathbb{T}_{\bar{N}}(u_N)\,
    |\bar{\Omega}^{(1)}(\l)
    \rangle|\Omega^{(2)}(\l)\rangle,\label{PF-1}
\eea where the monodromy matrix $\mathbb{T}_{\bar{i}}(u)$ is given by
\bea
 \mathbb{T}_{\bar{i}}(u)\equiv \mathbb{T}_{\bar{i}}(u;\xi_1,\ldots,\xi_N;\zeta)&=&
 \R_{\bar{i},N}(u_i\hspace{-0.1truecm}-\hspace{-0.1truecm}\xi_N)\ldots
    \R_{\bar{i},1}(u_i\hspace{-0.1truecm}-\hspace{-0.1truecm}\xi_1)\no\\
   &&\quad\quad\times K_{\bar{i}}(u_i)\,
    \R_{1,\bar{i}}(u_i\hspace{-0.1truecm}+\hspace{-0.1truecm}\xi_1)\ldots
    \R_{N,\bar{i}}(u_i\hspace{-0.1truecm}+\hspace{-0.1truecm}\xi_N).
    \label{Monodromy-1}
\eea The double-row matrix $\mathbb{T}(u)$ has played an important role in constructing the transfer matrix
for an open spin chain \cite{Skl88}. The QYBE (\ref{QYBE}) of the R-matrix and the reflection equation (\ref{RE-V}) of the
K-matrix  ensure that the monodromy matrix $\mathbb{T}_{\bar{i}}(u)$ satisfy the following exchange relation
\bea
\R_{\bar{i},\bar{j}}(u_i-u_j)\,\mathbb{T}_{\bar{i}}(u_i)\,\R_{\bar{j},\bar{i}}(u_i+u_j)\,
    \mathbb{T}_{\bar{j}}(u_j)=\mathbb{T}_{\bar{j}}(u_j)\,\R_{\bar{i},\bar{j}}(u_i+u_j)\,
    \mathbb{T}_{\bar{i}}(u_i)\,\R_{\bar{j},\bar{i}}(u_i-u_j).\label{RLL-1}
\eea

\subsection{The boundary states}

From the orthonormal basis $\{\e_i\}$ of $V$, we define
\bea
\hat{\imath}=\e_i-\overline{\e},~~\overline{\e}=
\frac{1}{2}\sum_{k=1}^{2}\e_k, \quad i=1,2,\qquad {\rm then}\,
\sum_{i=1}^2\hat{\imath}=0. \label{fundmental-vector} \eea Let
$\h$ be the Cartan subalgebra of $A_{1}$ and $\h^{*}$ be its dual.
A finite-dimensional diagonalizable  $\h$-module is a complex
finite-dimensional vector space $W$ with a weight decomposition
$W=\oplus_{\mu\in \h^*}W[\mu]$, so that $\h$ acts on $W[\mu]$ by
$x\,v=\mu(x)\,v$, $(x\in \h,\,v\in\,W[\mu])$. For example, the
non-zero weight spaces of the fundamental representation
$V_{\L_1}=\Cb^2=V$ are
\bea
 W[\hat{\imath}]=\Cb \e_i,~i=1,2.\label{Weight}
\eea

For a generic $m\in V$, define \bea m_i=\langle m,\e_i\rangle,
~~m_{ij}=m_i-m_j=\langle m,\e_i-\e_j\rangle,~~i,j=1,2.
\label{Def1}\eea Let $R(u,m)\in {\rm End}(V\otimes V)$ be the
R-matrix of the eight-vertex SOS model \cite{Bax82} given by
\bea
R(u;m)\hspace{-0.1cm}=\hspace{-0.1cm}
\sum_{i=1}^{2}R(u;m)^{ii}_{ii}E_{ii}\hspace{-0.1cm}\otimes\hspace{-0.1cm}
E_{ii}\hspace{-0.1cm}+\hspace{-0.1cm}\sum_{i\ne
j}^2\lt\{R(u;m)^{ij}_{ij}E_{ii}\hspace{-0.1cm}\otimes\hspace{-0.1cm}
E_{jj}\hspace{-0.1cm}+\hspace{-0.1cm}
R(u;m)^{ji}_{ij}E_{ji}\hspace{-0.1cm}\otimes\hspace{-0.1cm}
E_{ij}\rt\}, \label{R-matrix} \eea where $E_{ij}$ is the matrix
with elements $(E_{ij})^l_k=\d_{jk}\d_{il}$. The coefficient
functions are \bea
 &&R(u;m)^{ii}_{ii}=1,~~
   R(u;m)^{ij}_{ij}=\frac{\s(u)\,\s(m_{ij}-\eta)}
   {\s(u+\eta)\,\s(m_{ij})},~~i\neq j,\label{Elements1}\\[6pt]
 && R(u;m)^{ji}_{ij}=\frac{\s(\eta)\,\s(u+m_{ij})}
    {\s(u+\eta)\,\s(m_{ij})},~~i\neq j,\label{Elements2}
\eea  and $m_{ij}$ is defined in (\ref{Def1}). The R-matrix
satisfies the dynamical (modified) quantum Yang-Baxter equation
(or the star-triangle relation) \cite{Bax82}
\begin{eqnarray}
&&R_{1,2}(u_1-u_2;m-\eta h^{(3)})R_{1,3}(u_1-u_3;m)
R_{2,3}(u_2-u_3;m-\eta h^{(1)})\no\\
&&\qquad =R_{2,3}(u_2-u_3;m)R_{1,3}(u_1-u_3;m-\eta
h^{(2)})R_{1,2}(u_1-u_2;m).\label{MYBE}
\end{eqnarray}
Here we have adopted the convention
\bea R_{1,2}(u,m-\eta h^{(3)})\,v_1\otimes
v_2 \otimes v_3=\lt(R(u,m-\eta\mu)\otimes {\rm id }\rt)v_1\otimes
v_2 \otimes v_3,\quad {\rm if}\, v_3\in W[\mu]. \label{Action}
\eea Moreover, one may check that the R-matrix satisfies  the weight
conservation condition, \bea
  \lt[h^{(1)}+h^{(2)},\,R_{1,2}(u;m)\rt]=0,\label{Conservation}
\eea the unitary condition, \bea
 R_{1,2}(u;m)\,R_{2,1}(-u;m)={\rm id}\otimes {\rm
 id},\label{Unitary}
\eea and the crossing relation \bea
 R(u;m)^{kl}_{ij}=\varepsilon_{l}\,\varepsilon_{j}\,
   \frac{\s(u)\,\s((m-\eta\hat{\imath})_{21})}
   {\s(u+\eta)\,\s(m_{21})}\,R(-u-\eta;m-\eta\hat{\imath})
   ^{\bar{j}\,k}_{\bar{l}\,i},\label{Crossing}
\eea where
\bea \varepsilon_{1}=1,\,\varepsilon_{2}=-1,\quad {\rm
and}\,\, \bar{1}=2,\,\bar{2}=1.\label{Parity} \eea

Let us introduce two intertwiners which are
$2$-component  column vectors $\phi_{m,m-\eta\hat{\jmath}}(u)$
labelled by $\hat{1},\,\hat{2}$. The $k$-th element of
$\phi_{m,m-\eta\hat{\jmath}}(u)$ is given by \bea
\phi^{(k)}_{m,m-\eta\hat{\jmath}}(u)=\theta^{(k)}(u+2m_j),\label{Intvect}\eea
where the functions $\theta^{(j)}(u)$ are given in (\ref{Function-j}). Explicitly,
\bea \phi_{m,m-\eta\hat{1}}(u)=
\lt(\begin{array}{c}\theta^{(1)}(u+2m_1)\\[6pt]\theta^{(2)}(u+2m_1)\end{array}\rt),\qquad
\phi_{m,m-\eta\hat{2}}(u)=
\lt(\begin{array}{c}\theta^{(1)}(u+2m_2)\\[6pt]\theta^{(2)}(u+2m_2)\end{array}\rt).\eea
One can prove the following identity \cite{Fan98}
\bea
 {\rm det}\lt|\begin{array}{cc}\theta^{(1)}(u+2m_1)&\theta^{(1)}(u+2m_2)\\[6pt]
   \theta^{(2)}(u+2m_1)&\theta^{(2)}(u+2m_2)\end{array}\rt|=C(\tau)\,\s(u+m_1+m_2-\frac{1}{2})
   \,\s(m_{12}),
\eea where $C(\tau)$ is non-vanishing constant which depends on  $\tau$. This implies that
the two intertwiner vectors
$\phi_{m,m-\eta\hat{\imath}}(u)$ are linearly {\it independent}
for a generic $m\in V$.

Using the intertwiner vectors, one can derive the following face-vertex
correspondence relation \cite{Bax82}
\bea &&\R_{1,2}(u_1-u_2)
\phi^1_{m,m-\eta\hat{\imath}}(u_1)
\phi^2_{m-\eta\hat{\imath},m-\eta(\hat{\imath}+\hat{\jmath})}(u_2)
\no\\[6pt]&&~~~~~~= \sum_{k,l}R(u_1-u_2;m)^{kl}_{ij}
\phi^1_{m-\eta\hat{l},m-\eta(\hat{l}+\hat{k})}(u_1)
\phi^2_{m,m-\eta\hat{l}}(u_2). \label{Face-vertex} \eea  Then the
QYBE (\ref{QYBE}) of the vertex-type R-matrix $\R(u)$ is equivalent
to the dynamical Yang-Baxter equation (\ref{MYBE}) of the SOS
R-matrix $R(u,m)$. For a generic $m$, we can introduce other types
of intertwiners $\bar{\phi},~\tilde{\phi}$ which  are both row
vectors and satisfy the following conditions, \bea
  &&\bar{\phi}_{m,m-\eta\hat{\mu}}(u)
     \,\phi_{m,m-\eta\hat{\nu}}(u)=\d_{\mu\nu},\quad
     \tilde{\phi}_{m+\eta\hat{\mu},m}(u)
     \,\phi_{m+\eta\hat{\nu},m}(u)=\d_{\mu\nu},\label{Int2}\eea
{}from which one  can derive the relations,
\begin{eqnarray}
&&\sum_{\mu=1}^2\phi_{m,m-\eta\hat{\mu}}(u)\,
 \bar{\phi}_{m,m-\eta\hat{\mu}}(u)={\rm id},\label{Int3}\\[6pt]
&&\sum_{\mu=1}^2\phi_{m+\eta\hat{\mu},m}(u)\,
 \tilde{\phi}_{m+\eta\hat{\mu},m}(u)={\rm id}.\label{Int4}
\end{eqnarray}

Following the method in \cite{Fan98}, we check that the K-matrices $K(u)$ given by
(\ref{K-matrix}) can be expressed in terms
of the intertwiners and {\it diagonal\/} matrices $\K(\l|u)$
as follows
\bea &&K(u)^s_t=
\sum_{i,j}\phi^{(s)}_{\l-\eta(\hat{\imath}-\hat{\jmath}),
~\l-\eta\hat{\imath}}(u)
\K(\l|u)^j_i\bar{\phi}^{(t)}_{\l,~\l-\eta\hat{\imath}}(-u).
\label{K-F-1}
\eea Here the {\it
diagonal\/} matrix $\K(\l|u)$ is given
by \bea
&&\K(\l|u)\equiv{\rm Diag}(k(\l|u)_1,\,k(\l|u)_2)={\rm
Diag}(\frac{\s(\l_1+\zeta-u)}{\s(\l_1+\zeta+u)},\,
\frac{\s(\l_2+\zeta-u)}{\s(\l_2+\zeta+u)}).\label{K-F-3}
\eea
Although the vertex type K-matrix $K^{-}(u)$ given by
(\ref{K-matrix}) is  generally non-diagonal,
after the face-vertex transformation (\ref{K-F-1}), the face type
counterpart $\K(\l|u)$  becomes  diagonal. This
fact enabled the authors in \cite{Yan04-1,Yan07} to diagonalize the
transfer matrix of the open chains with non-diagonal terms
by applying the generalized algebraic Bethe
ansatz method developed in \cite{Yan04}.

Now we are in the position to construct the boundary states specifying the DW boundary
condition of the eight-vertex model with a non-diagonal reflection end, see figure 4. For
any vector $m\in V$, we introduce four states which live in the two N-tensor spaces of $V$
(one is indexed by $1,\ldots,N$ and the other is indexed by $\bar{1},\ldots,\bar{N}$) or their
dual spaces as follows:
\bea
 |\Omega^{(2)}(m)\rangle&=&
   \phi^{1}_{m,m-\eta\hat{2}}(\xi_1)\,
   \phi^{2}_{m-\eta\hat{2},m-2\eta\hat{2}}(\xi_2)\ldots
   \phi^{N}_{m-\eta (N-1)\hat{2},m-\eta N\hat{2}}(\xi_N),\label{Boundary-state-1}\\[8pt]
|\bar{\Omega}^{(1)}(m)\rangle&=&\phi^{\bar{1}}_{m-(N-2)\eta\hat{1},m-(N-2)\eta\hat{1}-\eta\hat{1}}(-u_1)
   \,\phi^{\bar{1}}_{m-(N-4)\eta\hat{1},m-(N-4)\eta\hat{1}-\eta\hat{1}}(-u_2)\no\\[8pt]
&&\times\ldots\phi^{\bar{N}}_{m+N\eta\hat{1},m+N\eta\hat{1}-\eta \hat{1}}(-u_N),
  \label{Boundary-state-2}\\[8pt]
\langle\Omega^{(1)}(m)|&=&
 \tilde{\phi}^{1}_{m,m-\eta\hat{1}}(\xi_1)\,
 \tilde{\phi}^{2}_{m-\eta\hat{1},m-2\eta\hat{1}}(\xi_2)\ldots
 \tilde{\phi}^{N}_{m-\eta(N-1)\hat{1},m-\eta N\hat{1}}(\xi_N),\label{Boundary-state-3}\\[8pt]
\langle\bar{\Omega}^{(2)}(m)|&=&
 \tilde{\phi}^{\bar{1}}_{m-N\eta\hat{1},m-N\eta\hat{1}-\eta\hat{2}}(u_1)\,
 \tilde{\phi}^{\bar{1}}_{m-(N-2)\eta\hat{1},m-(N-2)\eta\hat{1}-\eta\hat{2}}(u_2)\no\\[8pt]
&&\times \ldots\tilde{\phi}^{\bar{N}}_{m+\eta(N-2)\hat{1},m-\eta (N-2)\hat{1}-\eta\hat{2}}(u_N).\label{Boundary-state-4}
\eea The boundary states corresponding to  the DW
boundary condition  can be obtained through the above
states by special choices of $m$ and $i$ (for example,  $m$ is
specified to $\l$ which is related to the parameters of the K-matrix $K(u)$).
The DW partition
function $Z_N(\{u_{\a}\};\{\xi_i\};\l;\zeta)$ (\ref{PF}) of the eight-vertex model with a non-diagonal reflection end
becomes
\bea
 &&Z_N(\{u_{\a}\};\{\xi_i\};\l;\zeta)=\no\\[6pt]
 &&\quad\quad \tilde{\phi}^1_{\l,\l\hspace{-0.04truecm}-\hspace{-0.04truecm}\eta
  \hat{1}}(\xi_1)\ldots \tilde{\phi}^N_{\l\hspace{-0.04truecm}-\hspace{-0.04truecm}
  (N\hspace{-0.04truecm}-\hspace{-0.04truecm}1)\eta \hat{1},\l\hspace{-0.04truecm}-\hspace{-0.04truecm}N\eta\hat{1}}(\xi_N)
  \,\tilde{\phi}^{\bar{1}}_{\l\hspace{-0.04truecm}-\hspace{-0.04truecm}N\eta\hat{1},\l\hspace{-0.04truecm}-\hspace{-0.04truecm}N\eta
  \hat{1}\hspace{-0.04truecm}-\hspace{-0.04truecm}\eta\hat{2}}(u_1)\ldots \tilde{\phi}^{\bar{N}}
  _{\l\hspace{-0.04truecm}+\hspace{-0.04truecm}(N\hspace{-0.04truecm}-\hspace{-0.04truecm}2)\eta\hat{1},
  \l\hspace{-0.04truecm}+\hspace{-0.04truecm}(N\hspace{-0.04truecm}-\hspace{-0.04truecm}2)\eta\hat{1}
  \hspace{-0.04truecm}-\hspace{-0.04truecm}\eta\hat{2}}(u_N)\no\\[6pt]
 &&\quad\qquad\times R_{\bar{1},N}(u_1-\xi_N)\ldots R_{\bar{1},1}(u_1-\xi_1)\,
    K_{\bar{1}}(u_1)\,R_{1,\bar{1}}(u_1+\xi_1)\ldots R_{N,\bar{1}}(u_1+\xi_N)\no\\
 && \qquad\qquad\qquad\qquad \vdots\no\\
 &&\quad\qquad\times R_{\bar{N},N}(u_N-\xi_N)\ldots R_{\bar{N},1}(u_N-\xi_1)\,
    K_{\bar{N}}(u_N)\,R_{1,\bar{N}}(u_N+\xi_1)\ldots R_{N,\bar{N}}(u_N+\xi_N)
    \no\\[6pt]
 &&\quad\qquad\times \phi^{1}_{\l,\l\hspace{-0.04truecm}-\hspace{-0.04truecm}\eta\hat{2}}(\xi_1)\ldots
    \phi^{N}_{\l\hspace{-0.04truecm}-\hspace{-0.04truecm}(N\hspace{-0.04truecm}-\hspace{-0.04truecm}1)\eta\hat{2},
    \l\hspace{-0.04truecm}-\hspace{-0.04truecm}N\eta\hat{2}}(\xi_N)\,
    \phi^{\bar{1}}_{\l\hspace{-0.04truecm}-\hspace{-0.04truecm}(N\hspace{-0.04truecm}-\hspace{-0.04truecm}2)\eta\hat{1},
    \l\hspace{-0.04truecm}-\hspace{-0.04truecm}(N\hspace{-0.04truecm}-\hspace{-0.04truecm}1)\eta\hat{1}}
    (\hspace{-0.08truecm}-\hspace{-0.04truecm}u_1\hspace{-0.08truecm})
    \ldots\phi^{\bar{N}}_{\l\hspace{-0.04truecm}+\hspace{-0.04truecm}N\eta\hat{1},\l
    \hspace{-0.04truecm}+\hspace{-0.04truecm}(N\hspace{-0.04truecm}-\hspace{-0.04truecm}1)\eta\hat{1}}
    (\hspace{-0.08truecm}-\hspace{-0.04truecm}u_N\hspace{-0.08truecm}).\no\\
 && \label{PF-2}
\eea


\section{Partition function in terms of the face type monodromy matrix}
\label{Face} \setcounter{equation}{0}

Let us introduce the face type one-row monodromy matrix
\bea
 T_{F}(l|u)&\equiv &T^{F}_{0,1\ldots N}(l|u)\no\\[6pt]
 &=&R_{0,N}(u-\xi_N;l-\eta\sum_{i=1}^{N-1}h^{(i)})\ldots
    R_{0,2}(u-\xi_2;l-\eta h^{(1)})R_{0,1}(u-\xi_1;l),\no\\[6pt]
 &=&\lt(\begin{array}{ll}T_F(l|u)^1_1&T_F(l|u)^1_2\\T_F(l|u)^2_1&
   T_F(l|u)^2_2\end{array}\rt)
    \label{Monodromy-face-1}
\eea where $l$ is a generic vector in $V$. The monodromy matrix
satisfies the face type quadratic exchange relation
\cite{Fel96,Hou03}. Applying $T_F(l|u)^i_j$ to an arbitrary vector
$|i_1,\ldots,i_N\rangle$ in the N-tensor product space $V^{\otimes
N}$ given by \bea
   |i_1,\ldots,i_N\rangle=\e^1_{i_1}\ldots
   \e^N_{i_N},\label{Vector-V}
\eea we have \bea
 T_F(l|u)^i_j|i_1,\ldots,i_N\rangle&\equiv&
    T_F(m;l|u)^i_j|i_1,\ldots,i_N\rangle\no\\[6pt]
 &=&\sum_{\a_{N-1}\ldots\a_1}\sum_{i'_N\ldots i'_1}
 R(u-\xi_N;l-\eta\sum_{k=1}^{N-1}\hat{\imath}'_k)
   ^{i\,\,\,\,\,\,\,\,\,\,\,\,\,\,i'_N}_{\a_{N-1}\,i_N}\ldots\no\\[6pt]
 &&\quad\quad\times R(u-\xi_2;l-\eta\hat{\imath}'_1)^{\a_2\,i'_2}_{\a_1\,\,i_2}
 R(u-\xi_1;l)^{\a_1\,i'_1}_{j\,\,\,\,i_1}
   \,\,|i'_1,\ldots,i'_N\rangle,\label{Monodromy-face-2}
\eea where $m=l-\eta\sum_{k=1}^N\hat{\imath}_k$.

Now we compute the partition function $Z_N(\{u_{\a}\};\{\xi_i\};\l;\zeta)$ (\ref{PF}).
The expression (\ref{PF-2}) implies that
\bea
 Z_N(\{u_{\a}\};\{\xi_i\};\l;\zeta)
     \hspace{-0.38truecm}&=&\hspace{-0.38truecm}\langle\Omega^{(1)}(\l)|\,
     \tilde{\phi}^{\bar{1}}_{\l\hspace{-0.04truecm}-\hspace{-0.04truecm}(N-1)\eta\hat{1}+\eta\hat{2},\l\hspace{-0.04truecm}-\hspace{-0.04truecm}(N-1)\eta
     \hat{1}}(u_1)\, \mathbb{T}_{\bar{1}}(u_1)\,
     \phi^{\bar{1}}_{\l\hspace{-0.04truecm}-\hspace{-0.04truecm}(N-2)\eta\hat{1},\l\hspace{-0.04truecm}-\hspace{-0.04truecm}(N-1)\eta
     \hat{1}}(-u_1) \no\\
 &&\qquad\qquad\vdots\no\\
 &&\times \tilde{\phi}^{\bar{N}}_{\l\hspace{-0.04truecm}+\hspace{-0.04truecm}(N-1)\eta\hat{1}+\eta\hat{2},\l\hspace{-0.04truecm}+\hspace{-0.04truecm}(N-1)\eta
     \hat{1}}(u_N)\, \mathbb{T}_{\bar{N}}(u_N)\,
     \phi^{\bar{N}}_{\l\hspace{-0.04truecm}+\hspace{-0.04truecm}N\eta\hat{1},\l\hspace{-0.04truecm}+\hspace{-0.04truecm}(N-1)\eta
     \hat{1}}(-u_N)\,|\Omega^{(2)}(\l)\rangle.\no\\
\eea
With the help of the crossing
relation (\ref{Crossing}), the face-vertex
correspondence relation (\ref{Face-vertex}) and the relations (\ref{Int2}),
following the procedure in \cite{Yan04,Yan09}, we find that the partition function $Z_N(\{u_{\a}\};\{\xi_i\};\l;\zeta)$
can be expressed in terms of the face-type double-row monodromy operators
\bea
 &&Z_N(\{u_{\a}\};\{\xi_i\};\l;\zeta)=\langle 1,\ldots,1|\T^-_F(\l-2(M-1)\eta\hat{1},\l|u_1)^2_1
     \ldots\T^-_F(\l,\l|u_M)^2_1 \no\\[8pt]
     &&\qquad\qquad\times \T^-_F(\l+2\eta\hat{1},\l|u_{M+1})^2_1 \ldots
     \T^-_F(\l+N\eta\hat{1},\l|u_N)^2_1 |2,\ldots,2\rangle.\label{Scalar-3}
\eea  The above double-row monodromy  matrix operator $\T^-_F(m,\l|u)^2_1$ is given
by the one-row monodromy matrix operator $T_F(m;l|u)^i_j$
\cite{Yan11-1}
\bea
 &&\T^-_F(m,\l|u)^2_1=
 \frac{\s(m_{21})}{\s(\l_{21})}\prod_{k=1}^N
      \frac{\s(u+\xi_k)}{\s(u+\xi_k+\eta)}\no\\[6pt]
 &&\,\quad\times\lt\{
      \frac{\s(\l_1+\zeta-u)}{\s(\l_1+\zeta+u)}
      T_F(m,\l|u)^2_1
      T_F(m+\eta\hat{2},\l+\eta\hat{2}|-u-\eta)^2_2\rt.\no\\[6pt]
 &&\,\qquad-\lt.
      \frac{\s(\l_2+\zeta-u)}{\s(\l_2+\zeta+u)}
      T_F(m+2\eta\hat{2},\l|u)^2_2
      T_F(m+\eta\hat{1},\l+\eta\hat{1}|-u-\eta)^2_1\rt\}.\label{Expression-3}
 \eea


\section{ F-basis}
\label{F-basis} \setcounter{equation}{0}

In this section, after briefly reviewing the result \cite{Alb00-1} about the Drinfeld twist
\cite{Dri83} (factorizing F-matrix) of the eight-vertex SOS model, we
obtain the explicit expression of the double rows
monodromy operator $\T^-_F(m,\l|u)^2_1$ given by (\ref{Expression-3}) in the F-basis provided by the F-matrix.

\subsection{Factorizing Drinfeld twist $F$}
Let $ \mathcal{S}_N$ be the permutation group over indices
$1,\ldots,N$ and $\{s_i|i=1,\ldots,N-1\}$ be the set of
elementary permutations in $\mathcal{S}_N$. For each elementary
permutation $s_i$, we introduce the associated operator
$R^{s_i}_{1\ldots N}$ on the quantum space
\bea
  R^{s_i}_{1\ldots N}(l)\equiv R^{s_i}(l)=R_{i,i+1}
    (\xi_i-\xi_{i+1}|l-\eta\sum_{k=1}^{i-1}h^{(k)}),\label{Fundamental-R-operator}
\eea where $l$ is a generic vector in $V$. For any $s,\,s'\in
\mathcal{S}_N$, operator $R^{ss'}_{1\ldots N}$ associated with
$ss'$ satisfies the following composition law:
\bea
  R_{1\ldots N}^{ss'}(l)=R^{s'}_{s(1\ldots
  N)}(l)\,R^{s}_{1\ldots N}(l).\label{Rule}
\eea Let $s$ be decomposed in a minimal way in terms of
elementary permutations,
\bea
  s=s_{\b_1}\ldots s_{\b_p}, \label{decomposition}
\eea where $\b_i=1,\ldots, N-1$ and the positive integer $p$ is
the length of $s$. The composition law (\ref{Rule}) enables one
to obtain  operator $R^{s}_{1\ldots N}$ associated with each
$s\in\mathcal{S}_N $. The dynamical quantum Yang-Baxter equation
(\ref{MYBE}), weight conservation condition (\ref{Conservation})
and unitary condition (\ref{Unitary}) guarantee the uniqueness of
$R^{s}_{1\ldots N}$. Moreover, one may check that
$R^{s}_{1\ldots N}$ satisfies the following exchange relation
with the face type one-row monodromy matrix
(\ref{Monodromy-face-1}) \bea
  R^{s}_{1\ldots N}(l)T^F_{0,1\ldots N}(l|u)=T^F_{0,s(1\ldots N)}(l|u)
    R^{s}_{1\ldots N}(l-\eta h^{(0)}),\quad\quad \forall s\in
    \mathcal{S}_N.\label{Exchang-Face-1}
\eea

Now, we construct the face-type Drinfeld twist $F_{1\ldots
N}(l)\equiv F_{1\ldots N}(l;\xi_1,\ldots,\xi_N)$ \footnote{In this
paper, we adopt the convention: $F_{s(1\ldots N)}(l)\equiv
F_{s(1\ldots N)}(l;\xi_{s(1)},\ldots,\xi_{s(N)})$.} on the $N$-fold
tensor product space $V^{\otimes N}$, which  satisfies the
following three properties:
\bea
 &&{\rm I.\,\,\,\,lower-triangularity;}\\
 &&{\rm II.\,\,\, non-degeneracy;}\\
 &&{\rm III.\,factorizing \, property}:\,\,
 R^{s}_{1\ldots N}(l)\hspace{-0.08truecm}=\hspace{-0.08truecm}
    F^{-1}_{s(1\ldots N)}(l)F_{1\ldots N}(l), \,\,
 \forall s\in  \mathcal{S}_N.\label{Factorizing}
\eea Substituting (\ref{Factorizing}) into the exchange relation
(\ref{Exchang-Face-1}) yields the following relation
\bea
 F^{-1}_{s(1\ldots N)}(l)F_{1\ldots N}(l)T^F_{0,1\ldots N}(l|u)=
   T^F_{0,s(1\ldots N)}(l|u)F^{-1}_{s(1\ldots N)}(l-\eta h^{(0)})
   F_{1\ldots N}(l-\eta h^{(0)}).
\eea Equivalently,
\bea
 F_{1\ldots N}(l)T^F_{0,1\ldots N}(l|u)F^{-1}_{1\ldots N}(l-\eta h^{(0)})
   =F_{s(1\ldots N)}(l)T^F_{0,s(1\ldots N)}(l|u)
   F^{-1}_{s(1\ldots N)}(l-\eta h^{(0)}).\label{Invariant}
\eea Let us introduce the twisted monodromy matrix
$\tilde{T}^F_{0,1\ldots N}(l|u)$ by \bea
 \tilde{T}^F_{0,1\ldots N}(l|u)&=&
  F_{1\ldots N}(l)T^F_{0,1\ldots N}(l|u)F^{-1}_{1\ldots N}(l-\eta
  h^{(0)})\no\\[8pt]
  &=&\lt(\begin{array}{ll}\tilde{T}_F(l|u)^1_1&\tilde{T}_F(l|u)^1_2
  \\\tilde{T}_F(l|u)^2_1&
   \tilde{T}_F(l|u)^2_2\end{array}\rt).\label{Twisted-Mon-F}
\eea Then (\ref{Invariant}) implies that the twisted monodromy
matrix is symmetric under $\mathcal{S}_N$, namely, \bea
 \tilde{T}^F_{0,1\ldots N}(l|u)=\tilde{T}^F_{0,s(1\ldots
 N)}(l|u), \quad \forall s\in \mathcal{S}_N.
\eea

Define the F-matrix:
\bea
  F_{1\ldots N}(l)=\sum_{s\in
     \mathcal{S}_N}\sum^2_{\{\a_j\}=1}\hspace{-0.22truecm}{}^*\,\,\,\,
     \prod_{j=1}^NP^{s(j)}_{\a_{s(j)}}
     \,R^{s}_{1\ldots N}(l),\label{F-matrix}
\eea where $P^i_{\a}$ is the embedding of the project operator
$P_{\a}$ in the $i^{{\rm th}}$ space with matrix elements
$(P_{\a})_{kl}=\d_{kl}\d_{k\a}$. The sum $\sum^*$ in
(\ref{F-matrix}) is over all non-decreasing sequences of the
labels $\a_{s(i)}$:
\bea
  && \a_{s(i+1)}\geq \a_{s(i)}\quad {\rm if}\quad s(i+1)>s(i),\no\\[6pt]
  && \a_{s(i+1)}> \a_{s(i)}\quad {\rm if}\quad
  s(i+1)<s(i).\label{Condition}
\eea From (\ref{Condition}), $F_{1\ldots N}(l)$ obviously is a
lower-triangular matrix. Moreover, the F-matrix is non-degenerate
because  all its diagonal elements are non-zero. It was shown in \cite{Alb00-1}
that the F-matrix also satisfies the
factorizing property (\ref{Factorizing}).

\subsection{Completely symmetric  representations}
In the F-basis provided by the F-matrix (\ref{F-matrix}), the
twisted operators $\tilde{T}_F(l|u)^j_i$ defined by
(\ref{Twisted-Mon-F}) become  polarization free \cite{Alb00-1,Yan10}. Here
we present the results  relevant for our purpose
\bea
 &&\tilde{T}_F(l|u)^2_2=\frac{\s(l_{21}-\eta)}{\s\lt(l_{21}-\eta+
     \eta\langle H,\e_1\rangle\rt)}\otimes_{i}
     \lt(\begin{array}{ll}\frac{\s(u-\xi_i)}{\s(u-\xi_i+\eta)}&\\
     &1\end{array}\rt)_{(i)},\\[6pt]
 &&\tilde{T}_F(l|u)^2_1=\sum_{i=1}^N\frac{\s(\eta)
     \s(u\hspace{-0.08truecm}-\hspace{-0.08truecm}\xi_i
     \hspace{-0.08truecm}+\hspace{-0.08truecm}l_{12})}
     {\s(u\hspace{-0.08truecm}-\hspace{-0.08truecm}\xi_i
     \hspace{-0.08truecm}+\hspace{-0.08truecm}\eta)\s(l_{12})} E_{12}^i\otimes_{j\neq i}
     \lt(\begin{array}{ll}\frac{\s(u-\xi_j)\s(\xi_i-\xi_j+\eta)}{\s(u-\xi_j+\eta)\s(\xi_i-\xi_j)}&\\
     &1\end{array}\rt)_{(j)}.
\eea Applying  the above operators
to the arbitrary  state $|i_1,\ldots,i_N\rangle$ given by
(\ref{Vector-V}) leads to
\bea
 &&\tilde{T}_F(m,l|u)^2_2=\frac{\s(l_{21}-\eta)}{\s\lt(l_{2}-m_1-\eta\rt)}
     \otimes_{i}
     \lt(\begin{array}{ll}\frac{\s(u-\xi_i)}{\s(u-\xi_i+\eta)}&\\
     &1\end{array}\rt)_{(i)},\\[6pt]
 &&\tilde{T}_F(m,l|u)^2_1=\sum_{i=1}^N
     \frac{\s(\eta)
     \s(u-\xi_i+l_{12})}{\s(u-\xi_i+\eta)\s(l_{12})}\no\\[6pt]
 &&\quad\quad\quad\quad\quad\quad
     \times    E_{12}^i \otimes_{j\neq i}
     \lt(\begin{array}{ll}\frac{\s(u-\xi_j)\s(\
     \xi_i-\xi_j+\eta)}{\s(u-\xi_j+\eta)\s(\xi_i-\xi_j)}&\\
     &1\end{array}\rt)_{(j)}.
\eea It then follows that the  pseudo-particle creation
operator (\ref{Expression-3}) in the
F-basis  has the following completely symmetric
polarization free form:
\bea
 &&\tilde{\T}^-_F(m,\l|u)^2_1=\frac{\s( m_{12})}{\s(m_1-\l_2)}
   \prod_{k=1}^N\frac{\s(u+\xi_k)}{\s(u+\xi_k+\eta)}\no\\[6pt]
  &&\quad\quad\times \sum_{i=1}^N\frac{\s(\l_1+\zeta-\xi_i)\s(\l_2+\zeta+\xi_i)\s(2u) \s(\eta)}
   {\s(\l_1+\zeta+u)\s(\l_2+\zeta+u)\s(u-\xi_i+\eta)\s(u+\xi_i)}\no\\[6pt]
  &&\quad\quad\quad\quad\quad\quad \times
   E_{12}^i\otimes_{j\neq i}\lt(\begin{array}{ll}
   \frac{\s(u-\xi_j)\s(u+\xi_j+\eta)\s(\xi_i-\xi_j+\eta)}
   {\s(u-\xi_j+\eta)\s(u+\xi_j)\s(\xi_i-\xi_j)}&\\
   &1\end{array}\rt)_{(j)}.\label{Creation-operator-1}
\eea


\section{Determinant representation of the partition function}
\label{F} \setcounter{equation}{0}

In this section we  compute  the DW partition function $Z_N(\{u_{\a}\};\{\xi_i\};\l;\zeta)$ (\ref{Scalar-3})
using the expansion  of the twisted operator
$\tilde{\T}^-_F(m,\l|u)^2_1$ (\ref{Creation-operator-1}).

\subsection{Symmetric expression of the partition function}
From the definition of the F-matrix
$F_{1\ldots N}(l)$ (\ref{F-matrix}), we can show that the state
$|2,\ldots,2\rangle$ and the dual state
$\langle 1,\ldots,1|$  are invariant under the action of $F_{1\ldots N}(l)$, namely,
\bea
 F_{1\ldots N}(l)|2,\ldots,2\rangle&=&|2,\ldots,2\rangle,\label{Invariant-1}\\
 \langle 1,\ldots,1|F_{1\ldots N}(l)&=&\langle 1,\ldots,1|.
 \label{Invariant-2}
\eea Hence the DW partition function $Z_N(\{u_{\a}\};\{\xi_j\};\l;\zeta)$
can be expressed in terms of the twisted operator $\tilde{\T}^-_F(m,\l|u)^2_1$ as follow
\bea
 Z_N(\{u_{\a}\};\{\xi_i\};\l;\zeta)&=&\hspace{-0.36truecm}\langle 1,\ldots,1|\T^-_F(\l\hspace{-0.1truecm}-\hspace{-0.1truecm}
     (N\hspace{-0.1truecm}-\hspace{-0.1truecm}2)\eta\hat{1},\l|u_1)^2_1
     \ldots\T^-_F(\l\hspace{-0.1truecm}+\hspace{-0.1truecm}N\eta\hat{1},\l|u_N)^2_1 |2,\ldots,2\rangle\no\\[6pt]
 &=&\hspace{-0.36truecm}\langle 1,\ldots,1|\,F_{1\ldots N}(\l-N\eta\hat{1})\,\T^-_F(\l-(N-2)\eta\hat{1},\l|u_1)^2_1\ldots\no\\[6pt]
 &&\times \T^-_F(\l+N\eta\hat{1},\l|u_N)^2_1\, F^{-1}_{1\ldots N}(\l+N\eta\hat{1})\,|2,\ldots,2\rangle\no\\[6pt]
 &=&\hspace{-0.36truecm}\langle 1,\ldots,1|\tilde{\T}^-_F(\l\hspace{-0.1truecm}-\hspace{-0.1truecm}
     (N\hspace{-0.1truecm}-\hspace{-0.1truecm}2)\eta\hat{1},\l|u_1)^2_1
     \ldots\tilde{\T}^-_F(\l\hspace{-0.1truecm}+\hspace{-0.1truecm}N\eta\hat{1},\l|u_N)^2_1 |2,\ldots,2\rangle.\no
\eea Substituting the polarization free expression (\ref{Creation-operator-1}) of the twisted operator $\tilde{\T}^-_F(m,\l|u)^2_1$
into the above equation, we have {\small
\bea
&& \hspace{-0.46truecm}Z_N(\{u_{\a}\};\{\xi_i\};\l;\zeta)=
     \prod_{k=1}^M\frac{\s(\l_{12}+2k\eta)\s(\l_{12}-2k\eta+\eta)}
     {\s(\l_{12}+k\eta)\s(\l_{12}-k\eta+\eta)}
     \prod_{l=1}^N\prod_{i=1}^N\frac{\s(u_i+\xi_l)}{\s(u_i+\xi_l+\eta)}
     \langle 1,\ldots,1|\no\\[8pt]
  &&\hspace{-0.36truecm}\times\hspace{-0.1truecm}
     \sum_{i=1}^N\hspace{-0.1truecm}\frac{\s(\l_1\hspace{-0.1truecm}+\hspace{-0.1truecm}\zeta
     \hspace{-0.1truecm}-\hspace{-0.1truecm}\xi_i)\s(\l_2
     \hspace{-0.1truecm}+\hspace{-0.1truecm}\zeta\hspace{-0.1truecm}+\hspace{-0.1truecm}\xi_i)\s(2u_1) \s(\eta)}
     {\s(\hspace{-0.1truecm}\l_1\hspace{-0.1truecm}+\hspace{-0.1truecm}\zeta
     \hspace{-0.08truecm}+\hspace{-0.1truecm}u_1\hspace{-0.08truecm})\s(\hspace{-0.08truecm}\l_2
     \hspace{-0.1truecm}+\hspace{-0.1truecm}\zeta\hspace{-0.1truecm}+\hspace{-0.1truecm}u_1\hspace{-0.08truecm})
     \s(\hspace{-0.08truecm}u_1\hspace{-0.1truecm}-\hspace{-0.1truecm}\xi_i\hspace{-0.1truecm}+\hspace{-0.1truecm}\eta)
     \s(\hspace{-0.08truecm}u_1\hspace{-0.1truecm}+\hspace{-0.1truecm}\xi_i\hspace{-0.08truecm})}
     E_{12}^i\hspace{-0.1truecm}\otimes_{j\neq i}\hspace{-0.1truecm}\lt(\hspace{-0.28truecm}\begin{array}{ll}
     \frac{\s(u_1\hspace{-0.1truecm}-\xi_j)\s(u_1\hspace{-0.1truecm}+\xi_j\hspace{-0.1truecm}+\eta)\s(\xi_i\hspace{-0.1truecm}-\xi_j
     \hspace{-0.1truecm}+\eta)}
     {\s(u_1\hspace{-0.1truecm}-\xi_j\hspace{-0.1truecm}+\eta)\s(u_1\hspace{-0.1truecm}+\xi_j)\s(\xi_i\hspace{-0.1truecm}-\xi_j)}&\\
     &1\end{array}\hspace{-0.28truecm}\rt)_{\hspace{-0.18truecm}(j)}\no\\
  &&\qquad\qquad\vdots\no\\
  &&\hspace{-0.36truecm}\times\hspace{-0.1truecm}
     \sum_{i=1}^N\hspace{-0.1truecm}\frac{\s(\l_1\hspace{-0.1truecm}+\hspace{-0.1truecm}\zeta
     \hspace{-0.1truecm}-\hspace{-0.1truecm}\xi_i)\s(\l_2
     \hspace{-0.1truecm}+\hspace{-0.1truecm}\zeta\hspace{-0.1truecm}+\hspace{-0.1truecm}\xi_i)\s(2u_N) \s(\eta)}
     {\s(\hspace{-0.1truecm}\l_1\hspace{-0.1truecm}+\hspace{-0.1truecm}\zeta
     \hspace{-0.08truecm}+\hspace{-0.1truecm}u_N\hspace{-0.08truecm})\s(\hspace{-0.08truecm}\l_2
     \hspace{-0.1truecm}+\hspace{-0.1truecm}\zeta\hspace{-0.1truecm}+\hspace{-0.1truecm}u_N\hspace{-0.08truecm})
     \s(\hspace{-0.08truecm}u_N\hspace{-0.1truecm}-\hspace{-0.1truecm}\xi_i\hspace{-0.1truecm}+\hspace{-0.1truecm}\eta)
     \s(\hspace{-0.08truecm}u_N\hspace{-0.1truecm}+\hspace{-0.1truecm}\xi_i\hspace{-0.08truecm})}
     E_{12}^i\hspace{-0.1truecm}\otimes_{j\neq i}\hspace{-0.1truecm}\lt(\hspace{-0.28truecm}\begin{array}{ll}
     \frac{\s(u_N\hspace{-0.1truecm}-\xi_j)\s(u_N\hspace{-0.1truecm}+\xi_j\hspace{-0.1truecm}+\eta)\s(\xi_i\hspace{-0.1truecm}-\xi_j
     \hspace{-0.1truecm}+\eta)}
     {\s(u_N\hspace{-0.1truecm}-\xi_j\hspace{-0.1truecm}+\eta)\s(u_N\hspace{-0.1truecm}+\xi_j)\s(\xi_i\hspace{-0.1truecm}-\xi_j)}&\\
     &1\end{array}\hspace{-0.28truecm}\rt)_{\hspace{-0.18truecm}(j)}\no\\[8pt]
  &&\hspace{-0.36truecm}\times |2,\ldots,2\rangle.\no
\eea}Expanding the sum in the above equation corresponding to
the spectral parameter $u_N$ yields {\small
\bea
&& \hspace{-0.46truecm}Z_N(\{u_{\a}\};\{\xi_i\};\l;\zeta)=
     \prod_{k=1}^M\frac{\s(\l_{12}+2k\eta)\s(\l_{12}-2k\eta+\eta)}
     {\s(\l_{12}+k\eta)\s(\l_{12}-k\eta+\eta)}
     \prod_{l=1}^N\prod_{i=1}^N\frac{\s(u_i+\xi_l)}{\s(u_i+\xi_l+\eta)}\no\\[6pt]
  &&\hspace{-0.36truecm}\times\hspace{-0.1truecm}
     \sum_{i=1}^N\hspace{-0.1truecm}\frac{\s(\l_1\hspace{-0.1truecm}+\hspace{-0.1truecm}\zeta
     \hspace{-0.1truecm}-\hspace{-0.1truecm}\xi_i)\s(\l_2
     \hspace{-0.1truecm}+\hspace{-0.1truecm}\zeta\hspace{-0.1truecm}+\hspace{-0.1truecm}\xi_i)\s(2u_N) \s(\eta)}
     {\s(\hspace{-0.1truecm}\l_1\hspace{-0.1truecm}+\hspace{-0.1truecm}\zeta
     \hspace{-0.08truecm}+\hspace{-0.1truecm}u_N\hspace{-0.08truecm})\s(\hspace{-0.08truecm}\l_2
     \hspace{-0.1truecm}+\hspace{-0.1truecm}\zeta\hspace{-0.1truecm}+\hspace{-0.1truecm}u_N\hspace{-0.08truecm})
     \s(\hspace{-0.08truecm}u_N\hspace{-0.1truecm}-\hspace{-0.1truecm}\xi_i\hspace{-0.1truecm}+\hspace{-0.1truecm}\eta)
     \s(\hspace{-0.08truecm}u_N\hspace{-0.1truecm}+\hspace{-0.1truecm}\xi_i\hspace{-0.08truecm})}
     \prod_{l=1}^{N-1}\frac{\s(u_l-\xi_i)\s(u_l+\xi_i+\eta)}{\s(u_l-\xi_i+\eta)\s(u_l+\xi_i)}\no\\[6pt]
  &&\hspace{-0.16truecm}\times\hspace{-0.1truecm}\prod_{j\neq i}\frac{\s(\xi_j-\xi_i+\eta)}{\s(\xi_j-\xi_i)}
      \,\langle 1,\ldots,1|\no\\[8pt]
  &&\hspace{-0.16truecm}\times\hspace{-0.1truecm}
     \sum_{l\neq i}^N\hspace{-0.1truecm}\frac{\s(\l_1\hspace{-0.1truecm}+\hspace{-0.1truecm}\zeta
     \hspace{-0.1truecm}-\hspace{-0.1truecm}\xi_l)\s(\l_2
     \hspace{-0.1truecm}+\hspace{-0.1truecm}\zeta\hspace{-0.1truecm}+\hspace{-0.1truecm}\xi_l)\s(2u_1) \s(\eta)}
     {\s(\hspace{-0.1truecm}\l_1\hspace{-0.1truecm}+\hspace{-0.1truecm}\zeta
     \hspace{-0.08truecm}+\hspace{-0.1truecm}u_1\hspace{-0.08truecm})\s(\hspace{-0.08truecm}\l_2
     \hspace{-0.1truecm}+\hspace{-0.1truecm}\zeta\hspace{-0.1truecm}+\hspace{-0.1truecm}u_1\hspace{-0.08truecm})
     \s(\hspace{-0.08truecm}u_1\hspace{-0.1truecm}-\hspace{-0.1truecm}\xi_l\hspace{-0.1truecm}+\hspace{-0.1truecm}\eta)
     \s(\hspace{-0.08truecm}u_1\hspace{-0.1truecm}+\hspace{-0.1truecm}\xi_l\hspace{-0.08truecm})}
     E_{12}^i\hspace{-0.1truecm}\otimes_{j\neq l,i}\hspace{-0.1truecm}\lt(\hspace{-0.28truecm}\begin{array}{ll}
     \frac{\s(u_1\hspace{-0.1truecm}-\xi_j)\s(u_1\hspace{-0.1truecm}+\xi_j\hspace{-0.1truecm}+\eta)\s(\xi_l\hspace{-0.1truecm}-\xi_j
     \hspace{-0.1truecm}+\eta)}
     {\s(u_1\hspace{-0.1truecm}-\xi_j\hspace{-0.1truecm}+\eta)\s(u_1\hspace{-0.1truecm}+\xi_j)\s(\xi_l\hspace{-0.1truecm}-\xi_j)}&\\
     &1\end{array}\hspace{-0.28truecm}\rt)_{\hspace{-0.18truecm}(j)}\no\\
  &&\qquad\qquad\vdots\no\\
  &&\hspace{-0.36truecm}\times |2,\ldots,2\rangle.\no
\eea}Iterating the above procedure, we  obtain the
complete symmetric expression of the partition function
$Z_N(\{u_{\a}\};\{\xi_i\};\l;\zeta)$
\bea
Z_N(\{u_{\a}\};\{\xi_i\};\l;\zeta)&=&\prod_{k=1}^M\frac{\s(\l_{12}+2k\eta)\s(\l_{12}-2k\eta+\eta)}
  {\s(\l_{12}+k\eta)\s(\l_{12}-k\eta+\eta)}
  \prod_{l=1}^N\prod_{i=1}^N\frac{\s(u_i+\xi_l)}{\s(u_i+\xi_l+\eta)}\no\\[8pt]
  &&\quad \times {\cal Z}_N(\{u_{\a}\};\{\xi_i\};\l;\zeta),
  \label{partition-1}
\eea where the normalized partition function ${\cal Z}_N(\{u_{\a}\};\{\xi_i\};\l;\zeta)$ is
\bea
   {\cal Z}_N(\{u_{\a}\};\{\xi_i\};\l;\zeta)\hspace{-0.38truecm}&=&\hspace{-0.38truecm}
   \sum_{s\in\mathcal{S}_N}\prod_{n=1}^N\hspace{-0.08truecm}
   \lt\{\frac{\s(\l_1\hspace{-0.08truecm}+\hspace{-0.08truecm}\zeta
   \hspace{-0.08truecm}-\hspace{-0.08truecm}\xi_{i_{s(n)}})
   \s(\l_2\hspace{-0.08truecm}+\hspace{-0.08truecm}\zeta
   \hspace{-0.08truecm}+\hspace{-0.08truecm}\xi_{i_{s(n)}})
   \s(2u_n)\s(\eta)}
   {\s(\l_1\hspace{-0.08truecm}+\hspace{-0.08truecm}\zeta
   \hspace{-0.08truecm}+\hspace{-0.08truecm}u_n)
   \s(\l_2\hspace{-0.08truecm}+\hspace{-0.08truecm}\zeta
   \hspace{-0.08truecm}+\hspace{-0.08truecm}u_n)
   \s(u_n\hspace{-0.08truecm}-\hspace{-0.08truecm}\xi_{i_{s(n)}}
   \hspace{-0.08truecm}+\hspace{-0.08truecm}\eta)
   \s(u_n\hspace{-0.08truecm}+\hspace{-0.08truecm}\xi_{i_{s(n)}})}
   \rt.\no\\[8pt]
   && \times \prod_{k>n}^N\lt.
   \frac{\s(u_n-\xi_{i_{s(k)}}) \s(u_n+\xi_{i_{s(k)}}+\eta) \s(\xi_{i_{s(n)}}-\xi_{i_{s(k)}}+\eta)}
   {\s(u_n-\xi_{i_{s(k)}}+\eta) \s(u_n+\xi_{i_{s(k)}}) \s(\xi_{i_{s(n)}}-\xi_{i_{s(k)}})}
   \rt\}.\label{Function-B-2}
\eea

\subsection{Recursive relation and the determinant representation}
From the expression (\ref{Function-B-2}), it is easy to
check that the partition function  ${\cal Z}_N(\{u_{\a}\};\{\xi_i\};\l;\zeta)$ is a symmetric function of
$\{u_{\a}\}$ and $\{\xi_i\}$ separatively. Moreover,
we can show that the partition function ${\cal Z}_N(\{u_{\a}\};\{\xi_i\};\l;\zeta)$ satisfy the following recursive
relation
\bea
 \hspace{-1.2truecm}{\cal Z}_N(\{u_{\a}\};\{\xi_i\};\l;\zeta)&=&\sum_{i=1}^N
     \frac{\s(\l_1\hspace{-0.08truecm}+\hspace{-0.08truecm}\zeta
   \hspace{-0.08truecm}-\hspace{-0.08truecm}\xi_{i})
   \s(\l_2\hspace{-0.08truecm}+\hspace{-0.08truecm}\zeta
   \hspace{-0.08truecm}+\hspace{-0.08truecm}\xi_{i})
   \s(2u_N)\s(\eta)}
   {\s(\l_1\hspace{-0.08truecm}+\hspace{-0.08truecm}\zeta
   \hspace{-0.08truecm}+\hspace{-0.08truecm}u_N)
   \s(\l_2\hspace{-0.08truecm}+\hspace{-0.08truecm}\zeta
   \hspace{-0.08truecm}+\hspace{-0.08truecm}u_N)
   \s(u_N\hspace{-0.08truecm}-\hspace{-0.08truecm}\xi_{i}
   \hspace{-0.08truecm}+\hspace{-0.08truecm}\eta)
   \s(u_N\hspace{-0.08truecm}+\hspace{-0.08truecm}\xi_{i})}\no\\[8pt]
 &&\times \prod_{l=1}^{N-1}\frac{\s(u_l-\xi_i)\s(u_l+\xi_i+\eta)}
   {\s(u_l-\xi_i+\eta)\s(u_l+\xi_i)}\prod_{j\neq i}\frac{\s(\xi_j-\xi_i+\eta)}
   {\s(\xi_j-\xi_i)}\no\\[8pt]
 &&\times {\cal Z}_{N-1}(\{u_{\a}\}_{\a\neq N};\{\xi_j\}_{j\neq i};\l;\zeta).\label{Recursive-relation}
\eea
One can show that the initial condition
${\cal Z}_0(\{u_{\a}\};\{\xi_i\};\l;\zeta)=1$ and the  recursive relation (\ref{Recursive-relation})
{\it uniquely} determinate the partition function
${\cal Z}_N(\{u_{\a}\};\{\xi_i\};\l;\zeta)$ for any positive integer $N$. This fact allows us to
obtain the following determinant representation of the normalized partition function
${\cal Z}_N(\{u_{\a}\};\{\xi_i\};\l;\zeta)$:
\bea
{\cal Z}_N(\{u_{\a}\};\{\xi_i\};\l;\zeta)=
   \frac{\prod_{\a=1}^N\prod_{i=1}^N\s(u_{\a}-\xi_i)\s(u_{\a}+\xi_i+\eta)
   \,{\rm det}{\cal N}(\{u_{\a}\};\{\xi_i\})}
  {\prod_{\a>\b}\s(u_{\a}\hspace{-0.1truecm}-\hspace{-0.1truecm}
  u_{\b})\s(u_{\a}\hspace{-0.1truecm}+\hspace{-0.1truecm}u_{\b}
  \hspace{-0.1truecm}+\hspace{-0.1truecm}\eta)\prod_{k<l}
  \s(\xi_k\hspace{-0.1truecm}-\hspace{-0.1truecm}\xi_l)\s(\xi_k\hspace{-0.1truecm}+\hspace{-0.1truecm}\xi_l)},
  \label{partition-2}
\eea where the $N\times N$ matrix  ${\cal N}(\{u_{\a}\};\{\xi_i\})$ is given by
\bea
{\cal N}(\{u_{\a}\};\{\xi_i\})_{\a,j}&=&
  \frac{\s(\eta)\s(\l_1+\zeta-\xi_j)}
  {\s(u_{\a}-\xi_j)\s(u_{\a}+\xi_j+\eta)
  \s(\l_1+\zeta+u_{\a})}\no\\[8pt]
  &&\times \frac{\s(\l_2+\zeta+\xi_j)\s(2u_{\a})}
  {\s(\l_2+\zeta+u_{\a})
  \s(u_{\a}-\xi_j+\eta)
  \s(u_{\a}+\xi_j)}.\label{Matrix}
\eea

We shall prove the above determinant representation as follows. Firstly let us introduce two sets of functions
$\{B_I(\{u_{\a}\};\{\xi_i\};\l;\zeta)\,|\,I=1,\ldots, N\}$ and
$\{F_I(\{u_{\a}\};\{\xi_i\};\l;\zeta)\,|\,I=1,\ldots,N\}$ defined  respectively by
\bea
 \hspace{-1.2truecm} B_I(\{u_{\a}\};\{\xi_i\};\l;\zeta)&=&\hspace{-0.2truecm}
     \prod_{l=1}^I\frac{\s(\l_1\hspace{-0.08truecm}+\hspace{-0.08truecm}\zeta
     \hspace{-0.08truecm}+\hspace{-0.08truecm}u_l)\s(\l_2
     \hspace{-0.08truecm}+\hspace{-0.08truecm}\zeta\hspace{-0.08truecm}+\hspace{-0.08truecm}u_l)}
     {\s(\l_1\hspace{-0.08truecm}+\hspace{-0.08truecm}\zeta\hspace{-0.08truecm}-\hspace{-0.08truecm}\xi_l)
     \s(\l_2\hspace{-0.08truecm}+\hspace{-0.08truecm}\zeta\hspace{-0.08truecm}+\hspace{-0.08truecm}\xi_l)\s(2u_l)}\,
     {\cal Z}_I(\{u_{\a}\};\{\xi_i\};\l;\zeta),
     \label{B-function}\\[8pt]
  \hspace{-1.2truecm}F_I(\{u_{\a}\};\{\xi_i\};\l;\zeta)&=&\hspace{-0.28truecm}\frac{\prod_{\a=1}^I\prod_{j=1}^I
     \s(u_{\a}-\xi_j)\s(u_{\a}+\xi_j+\eta)}
     {\prod_{\a>\b}\s(u_{\a}\hspace{-0.1truecm}-\hspace{-0.1truecm}
     u_{\b})\s(u_{\a}\hspace{-0.1truecm}+\hspace{-0.1truecm}u_{\b}
     \hspace{-0.1truecm}+\hspace{-0.1truecm}\eta)\prod_{k<l}
     \s(\xi_k\hspace{-0.1truecm}-\hspace{-0.1truecm}\xi_l)\s(\xi_k\hspace{-0.1truecm}+\hspace{-0.1truecm}\xi_l)}
     \no\\[8pt]
  &&\hspace{-0.28truecm}\times {\rm det}\lt|\frac{\s(\eta)}
     {\s(u_{\a}\hspace{-0.1truecm}-\hspace{-0.1truecm}\xi_j)
     \s(u_{\a}\hspace{-0.1truecm}+\hspace{-0.1truecm}\xi_j\hspace{-0.1truecm}+\hspace{-0.1truecm}\eta)
     \s(u_{\a}\hspace{-0.1truecm}-\hspace{-0.1truecm}\xi_j\hspace{-0.1truecm}+\hspace{-0.1truecm}\eta)
     \s(u_{\a}\hspace{-0.1truecm}+\hspace{-0.1truecm}\xi_j)}\rt|.\label{F-function}
\eea Then the proof of (\ref{partition-2}) is equivalent to proving the equation
\bea
 B_I(\{u_{\a}\};\{\xi_i\};\l;\zeta)=F_I(\{u_{\a}\};\{\xi_i\};\l;\zeta),\quad {\rm for\,any\,positive\,integer\,\,I}.\label{Proof}
\eea We prove (\ref{Proof}) by induction.

\begin{itemize}

\item From direct calculation, we can show that (\ref{Proof}) holds for the case of $I=1$, namely,
\bea
 B_1(u_1;\xi_1;\l;\zeta)=F_1(u_1;\xi_1;\l;\zeta)=\frac{\s(\eta)}{\s(u_1-\xi_1+\eta)\s(u_1+\xi_1)}.\no
\eea

\item Suppose that (\ref{Proof}) holds for the  case of $I\leq N-1$. We now prove that it is satisfied also for
$I=N$. It is easy to check that both $B_N(\{u_{\a}\};\{\xi_i\};\l;\zeta)$ and $F_N(\{u_{\a}\};\{\xi_i\};\l;\zeta)$
are symmetric functions of $\{u_{\a}\}$. Hence it is sufficient to prove that they are equal  as
functions of $u_N$. The recursive relation (\ref{Recursive-relation})  of ${\cal Z}_N(\{u_{\a}\};\{\xi_i\};\l;\zeta)$
implies that $B_N(\{u_{\a}\};\{\xi_i\};\l;\zeta)$ satisfies the following
relation
\bea
 \hspace{-1.2truecm}B_N(\{u_{\a}\};\{\xi_i\};\l;\zeta)
    \hspace{-0.28truecm}&=&\hspace{-0.28truecm}\sum_{i=1}^N
    \hspace{-0.12truecm}\frac{\s(\eta)}
   {\s(u_N\hspace{-0.08truecm}-\hspace{-0.08truecm}\xi_{i}
   \hspace{-0.08truecm}+\hspace{-0.08truecm}\eta)
   \s(u_N\hspace{-0.08truecm}+\hspace{-0.08truecm}\xi_{i})}\hspace{-0.12truecm}
   \prod_{l=1}^{N-1}\frac{\s(u_l\hspace{-0.12truecm}-\hspace{-0.12truecm}\xi_i)
   \s(u_l\hspace{-0.12truecm}+\hspace{-0.12truecm}\xi_i\hspace{-0.12truecm}+\hspace{-0.12truecm}\eta)}
   {\s(u_l\hspace{-0.12truecm}-\hspace{-0.12truecm}\xi_i\hspace{-0.12truecm}+\hspace{-0.12truecm}\eta)
   \s(u_l\hspace{-0.12truecm}+\hspace{-0.12truecm}\xi_i)}\no\\
 &&\times \prod_{j\neq i}
   \frac{\s(\xi_j\hspace{-0.12truecm}-\hspace{-0.12truecm}\xi_i\hspace{-0.12truecm}+\hspace{-0.12truecm}\eta)}
   {\s(\xi_j\hspace{-0.12truecm}-\hspace{-0.12truecm}\xi_i)}\,
   B_{N-1}(\{u_{\a}\}_{\a\neq N};\{\xi_j\}_{j\neq i};\l;\zeta).\label{Recursive-relation-B}
\eea The determinant representation of the function $F_N(\{u_{\a}\};\{\xi_i\};\l;\zeta)$ implies that it satisfies
the following recursive relation
\bea
 \hspace{-1.2truecm}F_N(\hspace{-0.08truecm}\{u_{\a}\};\hspace{-0.08truecm}\{\xi_i\};\hspace{-0.08truecm}\l;\zeta)
    \hspace{-0.28truecm}&=&\hspace{-0.28truecm}\sum_{i=1}^N
    \hspace{-0.12truecm}\frac{\s(\eta)}
   {\s(u_N\hspace{-0.08truecm}-\hspace{-0.08truecm}\xi_{i}
   \hspace{-0.08truecm}+\hspace{-0.08truecm}\eta)
   \s(u_N\hspace{-0.08truecm}+\hspace{-0.08truecm}\xi_{i})}\hspace{-0.12truecm}
   \prod_{l=1}^{N-1}\frac{\s(u_l\hspace{-0.12truecm}-\hspace{-0.12truecm}\xi_i)
   \s(u_l\hspace{-0.12truecm}+\hspace{-0.12truecm}\xi_i\hspace{-0.12truecm}+\hspace{-0.12truecm}\eta)}
   {\s(u_N\hspace{-0.12truecm}-\hspace{-0.12truecm}u_l)
   \s(u_N\hspace{-0.12truecm}+\hspace{-0.12truecm}u_l+\eta)}\no\\[8pt]
 &&\times \prod_{j\neq i}\hspace{-0.08truecm}
   \frac{\s(u_N\hspace{-0.12truecm}-\hspace{-0.12truecm}\xi_j)
   \s(u_N\hspace{-0.12truecm}+\hspace{-0.12truecm}\xi_j\hspace{-0.12truecm}+\hspace{-0.12truecm}\eta)}
   {\s(\xi_j\hspace{-0.12truecm}-\hspace{-0.12truecm}\xi_i)\s(\xi_j\hspace{-0.12truecm}+\hspace{-0.12truecm}\xi_i)}\,
   \hspace{-0.08truecm}F_{N-1}(\hspace{-0.08truecm}\{u_{\a}\}_{\a\neq N};
   \hspace{-0.08truecm}\{\xi_j\}_{j\neq i};\hspace{-0.08truecm}\l;\zeta).\label{Recursive-relation-F}
\eea The determinant representation (\ref{F-function}) and its recursive relation (\ref{Recursive-relation-F}) of
$F_N(\{u_{\a}\};\{\xi_i\};\l;\zeta)$ and the recursive relation  (\ref{Recursive-relation-B}) of $B_N(\{u_{\a}\};\{\xi_i\};\l;\zeta)$
imply that these two functions, as functions of $u_N$, have the same simple poles located at \footnote{The determinant expression (\ref{F-function}) guarantees
that the apparent poles in (\ref{Recursive-relation-F}), which are located at $u_l,-u_l-\eta\,\,{\rm mod}(\Lambda_{\tau})$ for $l=1,\ldots,N-1$, do not
really be poles.}:
\bea
 \xi_i-\eta,\,-\xi_i\, \quad{\rm mod}(\Lambda_{\tau}),\quad\quad i=1,\ldots,N.\label{Poles}
\eea Direct calculation shows that the residues of the two functions at each simple pole (\ref{Poles}) are indeed
the same
\bea
  {\rm Res}_{u_N=z}\lt( F_N(\{u_{\a}\};\{\xi_i\};\l;\zeta)\rt)\hspace{-0.1truecm}=\hspace{-0.1truecm}
  {\rm Res}_{u_N=z}\lt( B_N(\{u_{\a}\};\{\xi_i\};\l;\zeta)\rt),
   \,z\hspace{-0.1truecm}=\hspace{-0.1truecm}\xi_i\hspace{-0.1truecm}-\hspace{-0.1truecm}\eta,\,
   \hspace{-0.1truecm}-\xi_i.\label{Res}
\eea  Let us introduce the function $f(u)$:
\bea
    f(u_N)=F_N(\{u_{\a}\};\{\xi_i\};\l;\zeta)-B_N(\{u_{\a}\};\{\xi_i\};\l;\zeta).
\eea The quasi-periodic properties of the $\s$-function,
\bea
 \s(u+1)=-\s(u),\quad\quad \s(u+\tau)=-e^{-2\sqrt{-1}\,\pi(u+\frac{\tau}{2})}\s(u),
\eea allow us to derive the following quasi-periodic properties of the function $f(u)$
\bea
  f(u+1)=f(u),\quad\quad f(u+\tau)=e^{-2\sqrt{-1}\,\pi(-2u-\eta-\tau)}f(u).
\eea This indicates \cite{Bax82} that the number of zero points minus that of poles of the function $f(u)$ inside the fundamental
domain generated by $1$ and $\tau$ should be $-2$. However
(\ref{Poles}) and (\ref{Res}) imply that the function $f(u)$ is actually analytic inside the fundamental domain. Therefore, we can conclude that
the function $f(u)$ must be zero function, i.e. $f(u)=0$. This means that (\ref{Proof})
holds for $I=N$.
\end{itemize}
Hence we have completed the proof of (\ref{partition-2}).

Finally from the expression (\ref{partition-1}), we obtain the determinant representation of the partition function
$Z_N(\{u_{\a}\};\{\xi_i\};\l;\zeta)$ (\ref{PF}) of the eight-vertex model
with a non-diagonal reflection end and  the DW boundary condition
\bea
 Z_N(\{u_{\a}\};\{\xi_i\};\l;\zeta)&=&\prod_{k=1}^M\frac{\s(\l_{12}+2k\eta)
     \s(\l_{12}-2k\eta+\eta)}{\s(\l_{12}+k\eta)\s(\l_{12}-k\eta+\eta)}
     \prod_{l=1}^N\prod_{i=1}^N\frac{\s(u_i+\xi_l)}{\s(u_i+\xi_l+\eta)}\no\\[8pt]
 &&\times \frac{\prod_{\a=1}^N\prod_{i=1}^N\s(u_{\a}-\xi_i)\s(u_{\a}+\xi_i+\eta)
   \,{\rm det}{\cal N}(\{u_{\a}\};\{\xi_i\})}
  {\prod_{\a>\b}\s(u_{\a}\hspace{-0.1truecm}-\hspace{-0.1truecm}
  u_{\b})\s(u_{\a}\hspace{-0.1truecm}+\hspace{-0.1truecm}u_{\b}
  \hspace{-0.1truecm}+\hspace{-0.1truecm}\eta)\prod_{k<l}
  \s(\xi_k\hspace{-0.1truecm}-\hspace{-0.1truecm}\xi_l)\s(\xi_k\hspace{-0.1truecm}+\hspace{-0.1truecm}\xi_l)},\no\\
  \label{partition-3}
\eea where the $N\times N$ matrix  ${\cal N}(\{u_{\a}\};\{\xi_i\})$ is given by (\ref{Matrix}).


\section{ Conclusions}
\label{C} \setcounter{equation}{0}

We have studied the partition function $Z_N(\{u_{\a}\};\{\xi_i\};\l;\zeta)$ of the eight-vertex model
with a non-diagonal reflection end defined by the K-matrix (\ref{K-matrix})  and the DW boundary
condition. The  DW boundary condition is specified by four boundary states
(\ref{Boundary-state-1})-(\ref{Boundary-state-4}). With the help of the F-basis provided by
the Drinfeld twist for the eight-vertex SOS model, we obtain the complete
symmetric expressions (\ref{partition-1})-(\ref{Function-B-2}) of the partition function. Such an explicit expression
allows us to derive the recursive relation (\ref{Recursive-relation}). Solving the recursive relation,
we find that the partition function $Z_N(\{u_{\a}\};\{\xi_i\};\l;\zeta)$ can be represented  as a single determinant (\ref{partition-3}),
contrary to the eight-vertex model without reflection end \cite{Pak08,Yan09-1,Ros09}.  Such a single determinant representation
will be crucial  for  constructing the explicit expressions of the Bethe vectors  and
further investigation of scalar productions of the open XYZ chain with non-diagonal boundary terms  \cite{Yan11-1}.

\section*{Acknowledgements}
The financial supports from  the National Natural Science
Foundation of China (Grant Nos. 11075126 and 11031005), Australian Research Council
and the NWU Graduate Cross-discipline Fund (08YJC24)
are gratefully acknowledged.



\begin{thebibliography}{99}
\bibitem{Kor82} V.\,E. Korepin, {\it Commun. Math. Phys.\/}
{\bf 86} (1982), 391.
\bibitem{Ize87} A.\,G. Izergin, {\it Sov. Phys. Dokl.} {\bf 32} (1987),
878.
\bibitem{Ize92} A.\,G. Izergin, D.A. Coker and V.E. Korepin,
{\it J. Phys.} {\bf A25} (1992), 4315.
\bibitem{Ess95} F.\,H.\,L. Essler, H. Frahm, A.\,G. Izergin and
V.\,E. Korepin,  {\it Commun. Math. Phys.} {\bf 174} (1995), 191.
\bibitem{Kor93} V.\,E. Korepin, N.\,M. Bogoliubov and A.\,G.
Izergin, {\it Quantum Inverse Scattering Method and Correlation
Functions}, Cambridge University Press, 1993.
\bibitem{Kit99} N. Kitanine, J.\,M. Maillet and V. Terras, {\it
     Nucl. Phys.\/} {\bf B 554} (1999), 647.
\bibitem{Ble05} P.M. Bleher and V.V. Fokin, Exact solution of the
six-vertex model with domain wall boundary conditions: Disordered
phase,  {\tt math-ph/0510033}.
\bibitem{Sog93} K. Sogo,  {\it J. Phys. Soc. Jpn.} {\bf 62} (1993)
1887.
\bibitem{Kup96} G. Kuperburg,  {\it Int. Math. Res. Not.} {\bf 3} (1996)
139; {\it Ann. of Math.\/} {\bf 156} (2002), 835.
\bibitem{Car06} A. Caradoc, O. Foda and N. Kitanine, {\it J. Stat.
Mech} (2006), P03012.
\bibitem{Fod08} O. Foda, M. Wheeler and M. Zuparic, {\it J.
Stat. Mech.} (2007), P10016;  {\it J. Stat. Mech.\/} (2008),
P02001.
\bibitem{Zha07} S.\,-Y. Zhao and Y.\,-Z. Zhang, {\it J. Math. Phys.\/} {\bf 48} (2007), 023504.
\bibitem{Pak08} S. Pakuliak, V. Rubtsov and A. Silantyev, {\it J. Phys.\/}
{\bf A 41} (2008), 295204.
\bibitem{Yan09-1} W.\,-L. Yang and Y.\,-Z. Zhang, {\it J. Math. Phys.\/} {\bf 50} (2009),
083518 [{\tt arXiv:0903.3089}].
\bibitem{Ros09} H. Rosengren, {\it Adv. in Appl. Math.\/} {\bf 43} (2009), 137.
\bibitem{Hao10} K. Hao, X. Chen, K.-J. Shi and W.\,-L Yang, {\it Chin. Phys. \/} {\bf B 20} (2011), 
010303.
\bibitem{Tsu98} O. Tsuchiya, {\it J. Math. Phys.} {\bf 39} (1998), 5946.
\bibitem{Skl88} E.\,K. Sklyanin, {\it J. Phys. \/} {\bf A 21}
(1988), 2375.
\bibitem{Wan02} Y.\,-S. Wang, {\it Nucl. Phys.\/} {\bf B 622}
(2002), 633.
\bibitem{Kit07} N. Kitanine, K.\,K. Kozlowski, J.\,M. Maillet, G.
     Niccoli, N.\, A. Slavnov and V. Terras, {\it J. Stat. Mech.\/}
     (2007), {\bf P10009}.
\bibitem{Dri83} V.\,G. Drinfeld, {\it Sov. Math. Dokl.\/} {\bf 28}
(1983), 667.
\bibitem{Mai00} J.\,M. Maillet and J. Sanchez de Santos, Drinfeld
twists and algebraic Bethe ansatz, {\it Amer. Math. Soc.
Transl.\/} {\bf 201} (2000), 137.
\bibitem{Nep04} R.\,I. Nepomechie,
{\it J. Stat. Phys.\/} {\bf 111} (2003), 1363; {\it J. Phys.\/}
{\bf A 37} (2004), 433.
\bibitem{Nep03} R.\,I. Nepomechie and F. Ravanini, {\it J.
Phys.\/} {\bf A 36} (2003), 11391; Addendum, {\it J. Phys. \/}
{\bf A 37} (2004), 1945.
\bibitem{Cao03} J. Cao, H.\,-Q. Lin, K.\,-J. Shi and Y. Wang, {\it
Nucl. Phys.\/} {\bf B 663} (2003), 487.
\bibitem{Yan04-1} W.\,-L. Yang, Y.\,-Z. Zhang and M. Gould, {\it
Nucl. Phys.\/} {\bf B 698} (2004), 312.
\bibitem{Gie05}J. de Gier and P. Pyatov, {\it J. Stat. Mech.\/}
(2004), {\bf P03002}; A. Nichols, V. Rittenberg and J. de Gier,
{\it J. Stat. Mech.\/} (2005), {\bf P05003}; J. de Gier, A.
Nichols, P. Pyatov and V. Rittenberg, {\it Nucl. Phys.\/} {\bf B
729} (2005), 387.
\bibitem{Yan04} W.\,-L. Yang and R. Sasaki, {\it Nucl. Phys.\/}
    {\bf B 679} (2004), 495; {\it J. Math. Phys.\/} {\bf 45} (2004),
    4301; W.\,-L. Yang, R. Sasaki and Y.\,-Z. Zhang, {\it JHEP}
    {\bf 09} (2004), 046.
\bibitem{Gal05} W. Galleas and M.\,J. Martins, {\it Phys. Lett.\/}
{\bf A 335} (2005), 167; C.\,S. Melo, G.\,A.\,P. Ribeiro and
M.\,J. Martins, {\it Nucl. Phys.\/} {\bf B 711} (2005), 565.
\bibitem{Gie05-1} J. de Gier and F.\,H.\,L. Essler, {\it Phys.
Rev. Lett.\/} {\bf 95} (2005), 240601; {\it J. Stat. Mech.\/}
(2006), {\bf P 12011}.
\bibitem{Baj06} Z. Bajnok, {\it J. Stat. Mech.\/} (2006), {\bf P06010}.
\bibitem{Yan05}  W.\,-L. Yang and  Y.\,-Z. Zhang,{\it JHEP} {\bf
12} (2004), 019; {\it JHEP} {\bf
01} (2005), 021; W.\,-L. Yang, Y.\,-Z. Zhang and R. Sasaki, {\it
Nucl. Phys.\/} {\bf 729} (2005), 594.
\bibitem{Doi06} A. Doikou and P.\,P. Martin, {\it J. Stat.
Mech.\/} (2006), {\bf P06004}; A. Dikou, {\it J. Stat. Mech.\/}
(2006), {\bf P09010}.
\bibitem{Mur06} R. Murgan, R.\,I. Nepomechie and C. Shi, {\it J.
Stat. Mech.\/}   (2006) {\bf P08006}; R. Murgan, {\it JHEP} {\bf 04} (2009), 076.
\bibitem{Bas07} P. Baseilhac and K. Koizumi, {\it J. Stat.
Mech.\/} (2007), {\bf P09006}.
\bibitem{Yan06} W.\,-L. Yang, R.\,I. Nepomechie and Y.\,-Z. Zhang,
{\it Phys. Lett.\/} {\bf B 633} (2006), 664; W.\,-L. Yang, and
Y.\,-Z. Zhang, {\it Nucl. Phys.\/} {\bf B 744} (2006), 312; L.
Frappat, R.\,I. Nepomechie and E. Ragoucy, {\it J. Stat. Mech.}
(2007), {\bf P09008}.
\bibitem{Gal08} W. Galleas, {\it Nucl. Phys.\/} {\bf B 790}
(2008), 524.
\bibitem{Yan07} W.\,-L. Yang and Y.\,-Z. Zhang, {\it JHEP} {\bf 04}
 (2007), 044; {\it Nucl. Phys.\/} {\bf B 789} (2008), 591.
\bibitem{Ami10} L. Amico, H. Frahm, A. Osterloh and T. Wirth,
 {\it Nucl. Phys. \/} {\bf B 839} (2010), 604.
\bibitem{Cra10} N. Crampe, E. Ragoucy and D. Simon, {\tt e-print: arXiv: 1009.4119}.
\bibitem{Fil10} G. Filali and N. Kitanine, {\it J. Stat. Mech.} {\bf 06}
(2010), L06001; {\tt e-print: arXiv: 1011.0660}.
\bibitem{Yan11} W.\,-L. Yang, X. Chen, J. Feng, K. Hao, B.\,-Y. Hou, K.\,-J. Shi
and Y.\,-Z. Zhang,  {\it Nucl. Phys.\/} {\bf B 844} (2011), 289.
\bibitem{Bax82} R.\,J. Baxter, {\it Exactly solved models in
statistical mechanics}, Academic Press, New York, 1982.
\bibitem{Ina94} T. Inami and H. Konno, {\it J. Phys.\/} {\bf A 27}
(1994), L913.
\bibitem{Hou95} B.\,Y. Hou, K.\,J. Shi, H. Fan and Z.\,-X. Yang,
{\it Commun. Theor. Phys.\/} {\bf 23} (1995), 163.
\bibitem{Whi50} E.\,T. Whittaker and G.\,N. Watson, {\it A course
of modern analysis}: {\bf 4th edn.}, Cambridge University Press,
2002.
\bibitem{Che84} I.\,V. Cherednik, {\it Theor. Math. Fiz\/} {\bf 61} (1984), 35.
\bibitem{Fan98} H. Fan, B.\,-Y. Hou, G.\,-L. Li and K.\,-J. Shi, {\it Phys. Lett.\/}
{\bf A 250} (1998), 79.
\bibitem{Fel96} G. Felder and A. Varchenko, {\it Nucl. Phys.\/}
  {\bf B 480} (1996), 485.
\bibitem{Hou03} B.\,Y. Hou, R. Sasaki and W.\,-L. Yang, {\it Nucl.
   Phys.\/} {\bf B 663} (2003), 467; {\it J. Math. Phys.\/} {\bf
   45} (2004), 559.
\bibitem{Yan09} W.\,-L. Yang and Y.\,-Z. Zhang, {\it Nucl. Phys. \/}
{\bf B 831} (2010), 408.
\bibitem{Yan11-1} W.\,-L. Yang, X. Chen, J. Feng, K. Hao, K.\,-J. Shi, Z.-Y. Yang
and Y.\,-Z. Zhang, {\tt arXiv:1101.2935}.
\bibitem{Alb00-1}  T.\,-D. Albert, H. Boos, R. Flume,
R.\,H. Poghossian and K. Rulig, {\it Lett. Math. Phys.\/} {\bf 53}
(2000), 201.
\bibitem{Yan10} W.\,-L. Yang, X. Chen, J. Feng, K. Hao, B.\,-Y. Hou, K.\,-J. Shi
and Y.\,-Z. Zhang, {\it JHEP} {\bf 01} (2011), 006.














\end{thebibliography}
\end{document}